\documentclass[twocolumn]{aastex631}

\usepackage{amsmath}
\usepackage{xcolor}
\usepackage{balance}
\usepackage{cuted}
\definecolor{MyBlue}{RGB}{0,0,255}
\usepackage{placeins} 
\usepackage{float}
\usepackage{booktabs}
\usepackage{natbib}
\usepackage{graphicx}
\usepackage{float}
\usepackage{hyperref}
\usepackage{subfigure}
\usepackage{longtable}
\usepackage{subfigure}
\usepackage{placeins}
\begin{document}


\title{\textit{NICER} Perspective on TeV Blazar Mrk~421: X-ray Variability and Particle Acceleration}

\author[0009-0000-4001-2012]{Sangeetha Kizhakkekalam}
\affiliation{Janusz Gil Institute of Astronomy, University of Zielona Góra,
ul. Szafrana 2, 65-516 Zielona Góra, Poland; email: \textcolor{blue}{g.bhatta@ia.uz.zgora.pl}}
\affiliation{Doctoral School of Exact and Technical Sciences, University of Zielona Góra,
al. Wojska Polskiego 69, 65-762 Zielona Góra, Poland; email: \textcolor{blue}{s.kizhakkekalam@stud.uz.zgora.pl}}

\author[0000-0002-0705-6619]{Gopal Bhatta}
\affiliation{Janusz Gil Institute of Astronomy, University of Zielona Góra,
ul. Szafrana 2, 65-516 Zielona Góra, Poland; email: \textcolor{blue}{g.bhatta@ia.uz.zgora.pl}}

\author[0000-0001-9092-2903]{Navaneeth P K}
\affiliation{Janusz Gil Institute of Astronomy, University of Zielona Góra,
ul. Szafrana 2, 65-516 Zielona Góra, Poland; email: \textcolor{blue}{g.bhatta@ia.uz.zgora.pl}}
\affiliation{Doctoral School of Exact and Technical Sciences, University of Zielona Góra,
al. Wojska Polskiego 69, 65-762 Zielona Góra, Poland; email: \textcolor{blue}{s.kizhakkekalam@stud.uz.zgora.pl}}

\author[0000-0003-4586-0744]{Tek P. Adhikari}
\affiliation{CAS Key Laboratory for Research in Galaxies and Cosmology, Department of Astronomy,
University of Science and Technology of China, Hefei, Anhui 230026, People’s Republic of China}
\affiliation{School of Astronomy and Space Science, University of Science and Technology of China,
Hefei, Anhui 230026, People’s Republic of China; email: \textcolor{blue}{tek@ustc.edu.cn}}




\begin{abstract}
Mrk~421 is one of the most fascinating blazars, widely studied across the electromagnetic spectrum using observations at various wavebands, from radio to the TeV gamma ray bands. We present the first detailed spectral and timing analysis of the TeV blazar Mrk~421 based on 45 X-ray observations from the \textit{NICER} X-ray telescope, collected over two years from 2022 to 2024. The source exhibits strong X-ray variability across intraday and long-term timescales. During this period, we observe a dramatic change in flux, from $\sim 50$ to $\sim 1380$~cts~s$^{-1}$, representing a $\sim 28$-fold increase. Spectral modeling with power-law, broken power-law, and log-parabolic functions shows that the log-parabola provides the most accurate description of the X-ray spectra. The hardness ratio analysis confirms a \textit{harder-when-brighter} trend, consistent with the anticorrelation between flux and photon index($\Gamma$). Correlation studies reveal a positive relation between the photon index ($\alpha$) and the curvature parameter ($\beta$) of the log-parabola model,  a negative correlation between $\beta$ and synchrotron peak energy ($E_{\mathrm{p}}$), and a positive correlation between $E_{\mathrm{p}}$ and flux. In addition, the observed rapid variability indicates that the X-ray emission originates from a compact region located close to the central engine. Furthermore, using a log-parabolic electron energy distribution within the synchrotron jet scenario, we simulate the observed anti-correlation between the  $E_{\rm p}$ and  $\beta$. These features can be interpreted within the framework of energy-dependent particle acceleration in blazar jets, which are often associated with turbulence, strong magnetic fields, and relativistic outflows.
\end{abstract}

\vspace*{1\baselineskip} 
\section{Introduction} \label{sec:intro}

Blazars, a subclass of radio-loud active galactic nuclei, are recognized for their relativistic jets oriented nearly along the observer’s line of sight \citep{urry1995unified}.~This alignment causes Doppler boosting of the jet’s non-thermal emission, making blazars particularly luminous and variable, especially at high energies \citep{bottcher2003coordinated, abdo2010spectral,bottcher2019,acciari2021}.~The sources exhibit significant variability throughout the electromagnetic spectrum, spanning timescales from minutes to years \citep{Navaneeth2025,bhattacharyya2020blazar,bhatta2020nature}. Their temporal variability, polarization behavior, and jet kinematics provide valuable insight into the physical processes operating in the vicinity of supermassive black holes \citep{ulrich1997variability, bhatta2016detection,bhatta2020nature}. The broadband spectral energy distribution (SED) of blazars is characterized by a two-peak structure \citep{Fossati1998,massaro2004log}. The first peak (low-energy peak), extending from the optical to X-ray band, is attributed to synchrotron radiation from relativistic electrons in the jet. In contrast, the second peak (high-energy peak) lies in the $\gamma$-ray regime and is observed at GeV–TeV energies, and its physical origin remains under active debate, with both leptonic and hadronic models proposed \citep{Maraschi1992,bottcher2003coordinated, Bottcher2007,bottcher2019}.~The origin of the seed photons for inverse-Compton scattering in blazars is described by two widely discussed leptonic models. In the synchrotron self-Compton (SSC) model, the same relativistic electrons that produce the synchrotron radiation also up-scatter these photons to higher energies \citep{Maraschi1992, Mastichiadis2002}. Alternatively, in the external Compton (EC) framework, the target photons originate from external regions of the AGN, such as the accretion disk \citep{Dermer1993}, the broad line region (BLR; \citealt{Sikora1994}), and the dusty torus \citep{Blazejowski2000}.~Based on the synchrotron peak in the SED and the optical spectral properties, the blazar population consists of two main subtypes: Flat-Spectrum Radio Quasars (FSRQs) and BL Lacertae objects (BL Lacs). FSRQs exhibit prominent broad emission lines with equivalent widths greater than 5\,\AA{} and are among the most powerful blazar sources, whereas BL Lacs, which display weak or absent emission lines with equivalent widths usually less than 5\,\AA{}, form a less powerful subclass.

In BL Lacs, the synchrotron peak lies at higher frequencies—from the IR to the X-ray regime—and they exhibit intense high-energy emission from hard X-rays to TeV gamma-rays, produced through synchrotron and inverse-Compton processes \citep{urry1995unified, ghisellini1998, falomo2014optical}. By contrast, FSRQs have their synchrotron peak at lower frequencies and are dominated by Compton emission  \citep{urry1995unified, Fossati1998, ghisellini1998, bhatta2025probing}. Based on the peak frequency ($\nu_{p}$) of the low-energy hump, BL Lac objects are further classified into low-energy-peaked BL Lacs (LBLs; $\nu_{p} < 10^{14}$ Hz), intermediate-energy-peaked BL Lacs (IBLs; $10^{14}$ Hz $< \nu_{p} < 10^{15}$ Hz), and high-energy-peaked BL Lacs (HBLs; $\nu_{p} > 10^{15}$ Hz) \citep[see][]{abdo2010spectral}.

In this paper, we present the results of a detailed spectral and timing analysis of the  Mrk~421. The layout of this paper is as follows. Section~\ref{sec:source} provides an overview of the source properties, and Section~\ref{sec:data} describes the \textit{NICER} data set and the reduction procedure. The analysis and results are presented in Section~\ref{sec:analysis}, followed by a discussion and comparison with previous studies in Section~\ref{sec:discussion}. Our conclusions are summarized in Section~\ref{sec:conclusion}.
\section{Source Mrk~421}\label{sec:source}
Markarian~421 (Mrk~421), located at RA = 11\textsuperscript{h}04\textsuperscript{m}27.31\textsuperscript{s} and Dec = +38$^\circ$12$'$31.8$''$ (J2000), is a prototypical high-energy-peaked BL Lac (HBL) object.
Mrk~421 was the first blazar---and indeed the first extragalactic source---discovered to be a TeV emitter: initially detected as a weak $\gamma$-ray source by EGRET \citep{lin1992detection}, it was followed nine months later by a clear 0.5--1.5~TeV detection with the Whipple telescope \citep{punch1992detection, Petry1996}.
With a redshift of $z = 0.031$, Mrk~421 is one of the nearest TeV blazars, where minimal attenuation by the extragalactic background light (EBL) allows more accurate constraints on its intrinsic VHE spectrum \citep{punch1992detection, Aharonian2006, Mazin2007}.

The source has been the focus of numerous multiwavelength (MWL) campaigns with diverse temporal and spectral coverage, involving both space- and ground-based observatories, and is persistently bright in the optical, X-ray, and TeV bands, often exhibiting rapid ($\lesssim$1--2~day) variability in the X-ray and TeV regimes \citep{blazejowski2005, Aleksic2015, Balokovi2016, Markowitz2022}.

Optical studies of Mrk~421 reveal significant long-term and intraday variability, as shown in early monitoring campaigns \citep{Tosti1998,Gaidos1996}. More recent photometric and polarimetric analyses have established a clear \textit{bluer-when-brighter} trend and strong polarization variability, including EVPA rotations, indicating a dynamically evolving magnetic-field structure within the jet \citep{Chen2014,Carnerero2017,Fraija2017,Singh2019}.

Observations by the Large High Altitude Air Shower Observatory (\textit{LHAASO}) have revealed detections of the source in the PeV energy range, placing Mrk~421 among the most extreme particle accelerators \citep{cao2024first, abeysekara2020great}. The major outburst in 2010 provided valuable information on the multiwavelength behavior of the source; strong temporal correlations were observed between the X-ray and VHE $\gamma$-ray bands \citep{DiGesu2023,Aleksic2015}. In the same period, \textit{VERITAS} recorded a historic VHE $\gamma$-ray flare from Mrk~421, reaching a flux of approximately 27 Crab Units (CU) above 1~TeV \citep{abeysekara2020great,abe2025characterization}.

Recent polarimetric observations have documented significant short-term variability in X-ray polarization, including a $\sim$90$^\circ$ rotation in the polarization angle around the jet axis. Two \textit{IXPE} observations of Mrk~421 analyzed by \citet{Kim2024} showed a consistently high degree of X-ray polarization ($\sim$14\%) together with intermittent variations in the polarization angle. Moreover, Mrk~421 exhibits pronounced X-ray variability, as revealed by observations from \textit{XMM-Newton}, \textit{Swift}, \textit{NuSTAR}, and \textit{AstroSat}.~These studies have shown strong correlations between the soft and hard X-ray bands, with the spectra generally described by power-law or curved model \citep{kapanadze2016, akbar2025probing, Noel2022, Brinkmann2001, Bhatta2025Nustar}. 

Mrk~421 shows a characteristic curved X-ray spectrum, typical of high energy peaked BL lacs \citep{fossati2000x}, in which flux variability is frequently linked with pronounced spectral evoltion \citep{tramacere2007signatures,massaro2004}. Complementary observations with \textit{Chandra} have detected absorption features attributable to warm–hot intergalactic medium (WHIM) filaments along the line of sight, providing important evidence for their role in tracing the missing baryons of the cosmic web \citep{Brinkmann2001,Kaastra2006}. Stochastic and time-domain behavior in Mrk~421 has also been examined in detail by \citet{bhattacharyya2020blazar}, \citet{Goyal2018}, \citet{Sobolewska2014}, \citet{Panis2025}. Investigating time-resolved X-ray emission using different spectral models offers an effective means of identifying correlations between key model parameters. \citet{Hota2021} conducted such an analysis, exploring the temporal evolution of X-ray spectra under several modeling approaches.

\vspace{-\baselineskip}

\begin{deluxetable*}{cccccc}

\tablecaption{Observational properties of the \textit{NICER} observations of the TeV blazar Mrk~421. 
Column 1: serial number; Column 2: observation ID; Column 3: exposure time (s); 
Column 4: observation date; Column 5: average count rate with error; 
Column 6: fractional variability with error.\label{tab:obs-details}}
\tablehead{
\colhead{SN} & 
\colhead{ObsID} & 
\colhead{Exposure (s)} & 
\colhead{Date} & 
\colhead{Av.~CR $\pm$ Err} & 
\colhead{Fvar $\pm$ Err (\%)}
}
\colnumbers
\startdata
1 & 5100110101 & 2056.00 & 2022-03-25 & 93.34 ± 1.82 & 6.11 ± 0.24 \\
2 & 5100110102 & 3102.00 & 2022-03-27 & 83.73 ± 1.24 & 6.20 ± 0.22 \\
3 & 5100110104 & 1491.00 & 2022-03-28 & 50.62 ± 0.56 & 3.00 ± 0.48 \\
4 & 6704012801 & 1719.00 & 2023-04-08 & 439.53 ± 1.32 & 0.75 ± 0.15 \\
5 & 6704013201 & 1085.00 & 2023-04-14 & 729.71 ± 2.50 & 0.72 ± 0.12 \\
6 & 6704013801 & 1223.00 & 2023-04-23 & 706.94 ± 1.18 & 0.28 ± 0.11 \\
7 & 6704016701 & 1621.00 & 2023-05-17 & 597.48 ± 0.79 & 0.21 ± 0.11 \\
8 & 6100110102 & 7195.00 & 2023-05-18 & 936.88 ± 12.25 & 7.84 ± 0.04 \\
9 & 6100110103 & 5024.00 & 2023-05-20 & 826.62 ± 5.09 & 3.13 ± 0.06 \\
10 & 6704016901 & 1925.00 & 2023-05-23 & 857.42 ± 1.04 & 0.26 ± 0.08 \\
11 & 6704017001 & 2168.56 & 2023-05-23 & 745.76 ± 1.35 & 0.48 ± 0.08 \\
12 & 6100110101 & 1603.00 & 2023-05-24 & 1043.25 ± 1.06 & 0.15 ± 0.09 \\
13 & 6100110104 & 2359.00 & 2023-05-24 & 725.67 ± 16.10 & 7.99 ± 0.09 \\
14 & 6704017101 & 2173.00 & 2023-05-24 & 820.06 ± 1.78 & 0.61 ± 0.08 \\
15 & 6704017201 & 1134.00 & 2023-05-24 & 1372.59 ± 1.84 & 0.23 ± 0.08 \\
16 & 6704018501 & 1476.53 & 2023-05-31 & 652.43 ± 7.72 & 1.97 ± 0.33 \\
17 & 6704020601 & 1762.00 & 2023-06-23 & 470.65 ± 8.86 & 6.23 ± 0.13 \\
18 & 6100110110 & 1707.25 & 2023-11-16 &  1109.40 ± --& -- \\
19 & 6100110111 & 1868.54 & 2023-11-16 & 1380.38 ± 24.84 & 6.47 ± 0.13 \\
20 & 6100110116 & 3387.71 & 2023-11-23 & 782.21 ± 6.89 & 3.67 ± 0.26 \\
21 & 6100110132 & 16291.24 & 2024-01-01 & 1063.46 ± 2.39 & 0.28 ± 0.16 \\
22 & 6100110133 & 12038.23 & 2024-01-01 & 943.76 ± 17.97 & 3.79 ± 0.17 \\
23 & 6100110134 & 18990.48 & 2024-01-01 & 1030.55 ± 19.21 & 7.90 ± 0.09 \\
24 & 6704028701 & 1111.00 & 2024-01-01 & 1073.45 ± 1.10 & 0.08 ± 0.11 \\
25 & 6100110135 & 11074.95 & 2024-01-02 & 1071.52 ± 10.30 & 4.29 ± 0.07 \\
26 & 6100110136 & 5960.00 & 2024-01-02 & 979.43 ± 9.62 & 2.57 ± 0.13 \\
27 & 6100110137 & 4171.00 & 2024-01-02 & 977.63 ± 16.23 & 7.60 ± 0.06 \\
28 & 6100110141 & 1295.00 & 2024-01-04 & 734.57 ± 2.12 & 0.70 ± 0.12 \\
29 & 6704029501 & 1009.00 & 2024-01-07 & 805.22 ± 0.80 & -- \\
30 & 6704030801 & 1089.00 & 2024-01-17 & 534.11 ± 1.37 & 0.48 ± 0.14 \\
31 & 6100110143 & 1773.00 & 2024-01-30 & 419.77 ± 0.53 & 0.06 ± 0.17 \\
32 & 6704031701 & 1588.54 & 2024-01-30 & 705.36 ± 1.73 & 0.59 ± 0.10 \\
33 & 6704031901 & 1678.00 & 2024-01-30 & 576.52 ± 2.66 & 1.19 ± 0.10 \\
34 & 6704032101 & 1747.00 & 2024-01-30 & 468.85 ± 1.32 & 0.71 ± 0.12 \\
35 & 6704032301 & 1399.00 & 2024-02-01 & 314.29 ± 0.52 & 0.15 ± 0.17 \\
36 & 6704032801 & 1072.00 & 2024-02-09 & 529.67 ± 1.04 & 0.30 ± 0.15 \\
37 & 6704033101 & 1115.00 & 2024-02-12 & 1048.44 ± 1.63 & 0.27 ± 0.10 \\
38 & 6704035901 & 1370.23 & 2024-03-16 & 433.77 ± 0.83 & 0.21 ± 0.18 \\
39 & 6704035701 & 1246.24 & 2024-03-19 & 281.62 ± 0.61 & 0.11 ± 0.29 \\
40 & 7566013301 & 1304.00 & 2024-04-12 & 387.77 ± 0.77 & -- \\
41 & 7100110102 & 4435.62 & 2024-04-14 & 410.30 ± 8.98 & 6.55 ± 0.18 \\
42 & 7566015901 & 1365.54 & 2024-05-09 & 223.64 ± 1.14 & 1.12 ± 0.23 \\
43 & 7566016701 & 1179.00 & 2024-05-17 & 258.24 ± 0.35 & -- \\
44 & 7566039501 & 1139.00 & 2024-12-30 & 232.65 ± 0.52 & 0.23 ± 0.22 \\
45 & 6704039501 & 1201.00 & 2024-12-31 & 229.85 ± 0.51 & 0.22 ± 0.24 \\
\enddata
\footnotesize A dashed entry (``--'') indicates that the total variance of the observation is smaller than the mean square of the measurement uncertainties.

\end{deluxetable*}
\vspace{-\baselineskip}

\section{OBSERVATION AND DATA PROCESSING} \label{sec:data}

For our analysis, we utilized archival data from the Neutron Star Interior Composition Explorer (\textit{NICER}), a soft X-ray timing instrument aboard the International Space Station (ISS) \citep{Gendreau2016, Arzoumanian2014}. \textit{NICER} provides high-sensitivity rotation-resolved spectroscopy of thermal and non-thermal X-ray emissions from various astronomical sources primarily neutron stars in the 0.2–12 keV energy range. As a non-imaging X-ray telescope, \textit{NICER} is designed primarily for high-precision timing and spectral studies \citep{Gendreau2016}. The instrument consists of 56 aligned Focal Plane Modules (FPMs), each integrating an X-ray concentrator (XRC) optic and a silicon drift detector (SDD), of which 52 are currently operational \citep{Malacaria2023}. An individual Focal Plane Module (FPM) comprises a detector, a preamplifier, and a thermoelectric cooler. Each XRC collects X-rays over a large geometric area from a roughly 30 arcmin$^2$ region of the sky and focuses them onto a small SDD\footnote{\url{https://heasarc.gsfc.nasa.gov/docs/nicer/}}.
The SDD detects individual photons, records their energies with good spectral resolution (a few per cent) in the energy band of 0.2–12 keV and provides a time resolution of approximately 100 nanoseconds and an energy resolution of about 150 eV at 6.5 keV. With its exceptional temporal resolution (100~ns) and minimal instrumental background, \textit{NICER} provides valuable capabilities for diverse high-energy astrophysical applications, extending beyond its primary neutron-star science goals to studies of relativistic jet dynamics and nonthermal emission processes in blazars. The sensitivity of \textit{NICER} in the soft X-ray band provides a substantial advantage for investigating high-synchrotron-peaked blazars such as Mrk 421, where the synchrotron emission extends into the X-ray regime.

In this study, we analyzed \textit{NICER} observations of Mrk~421, a well-known high-synchrotron-peaked (HSP) blazar and bright TeV emitter, which is an excellent candidate for X-ray variability and spectral index evolution studies. The observational details are summarized in Table~\ref{tab:obs-details}, where we present the record of all \textit{NICER} observations used in this work. For the analysis, we restricted the energy range to 0.4–10~keV to ensure optimal calibration, minimize interstellar absorption at lower energies, and reduce background contamination at higher energies.

Standard \textit{NICER} data reduction procedures were performed using HEASoft v6.34 with the CALDB release 20240206 from NASA’s HEASARC archive and the most recent NICERDAS calibration files, in combination with XSPEC v12.14.1. Good time intervals (GTIs) were generated by excluding periods of elevated background, such as passages of the South Atlantic Anomaly (SAA), low Earth elevation, and times of enhanced particle activity. Barycentric corrections of the \textit{NICER} event files were applied using the \texttt{barycorr} tool from FTOOLS, adopting the JPL DE405 planetary ephemeris and the ICRS reference frame, with source coordinates \citep{Ng2021}. Finally, the task \texttt{nicerl2} was applied to obtain the calibrated and filtered clean event files required for further spectral and timing analysis.

For spectral analysis, the standard task \texttt{nicerl3-spect} was used to generate the spectral products, including the source spectrum and the response files (ARF and RMF). We followed the NICERV5 Level 2 processing screening criteria, which exclude intervals affected by poor pointing, Earth-limb proximity, solar contamination, and enhanced particle background. The \texttt{SCORPEON} background model was employed to generate background spectra for the \textit{NICER} observations, accounting for contributions from non-X-ray background (NXB; e.g., trapped electrons, the South Atlantic anomaly, cosmic rays) and the astrophysical X-ray background (CXB; diffuse local or Galactic emission, neutral oxygen, and solar wind charge exchange). Unlike traditional background models based on static libraries, \texttt{SCORPEON} is parameterized and can be fitted alongside the source model as a function of both time and sky location.

\clearpage

\begin{figure*}[!t]
\centering

\includegraphics[width=0.95\textwidth]{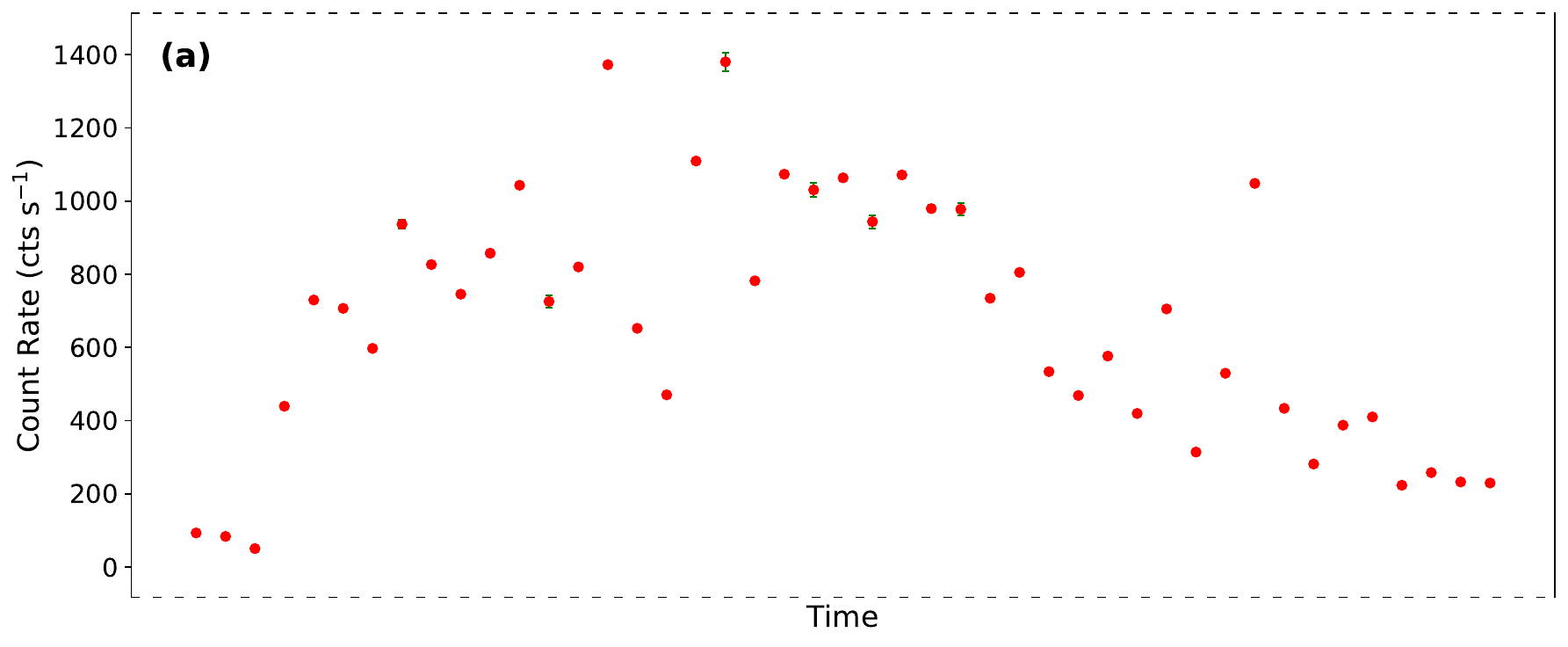}

\vspace{0.3em}

\includegraphics[width=0.44\textwidth]{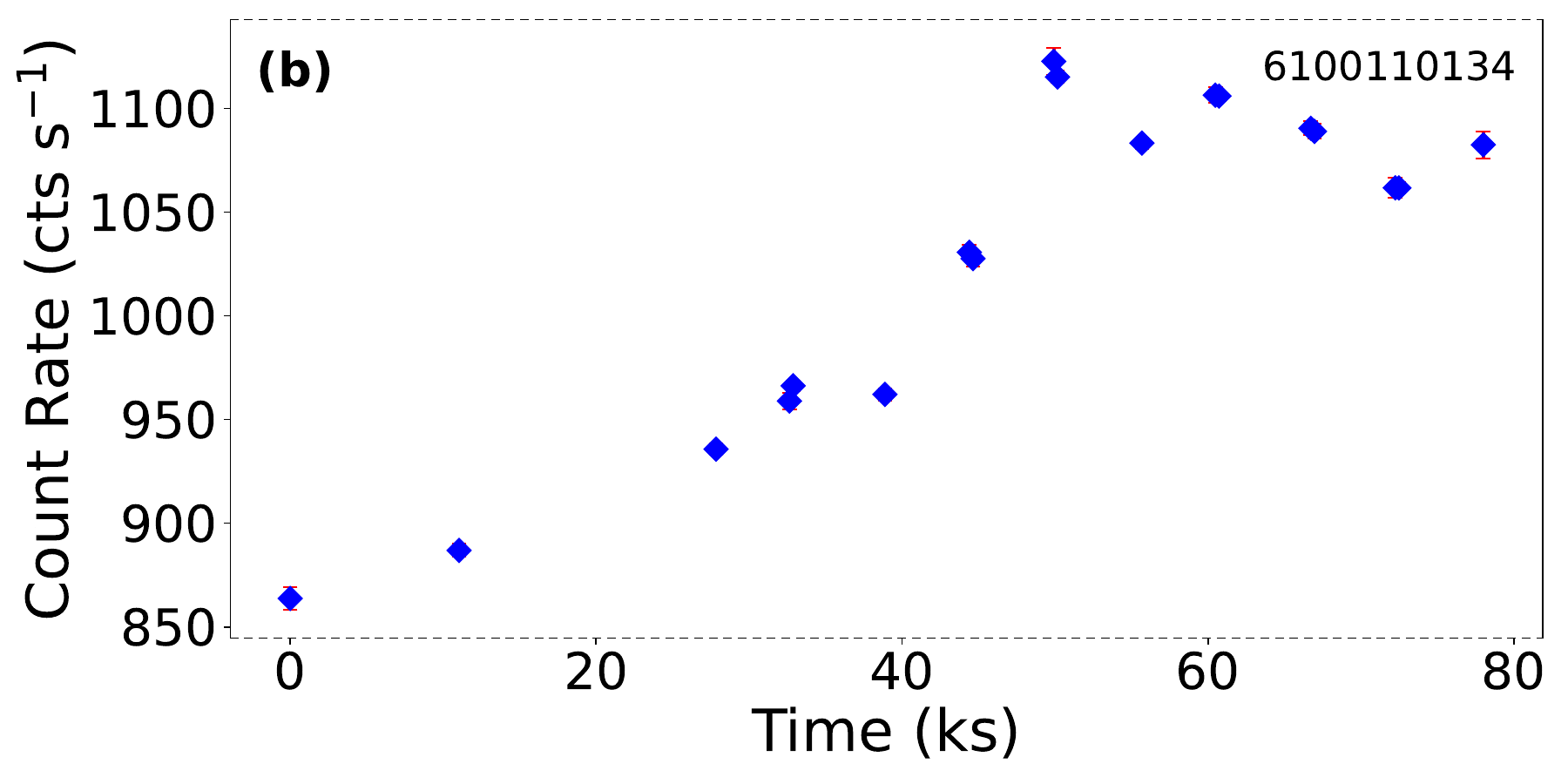}\hfill
\includegraphics[width=0.44\textwidth]{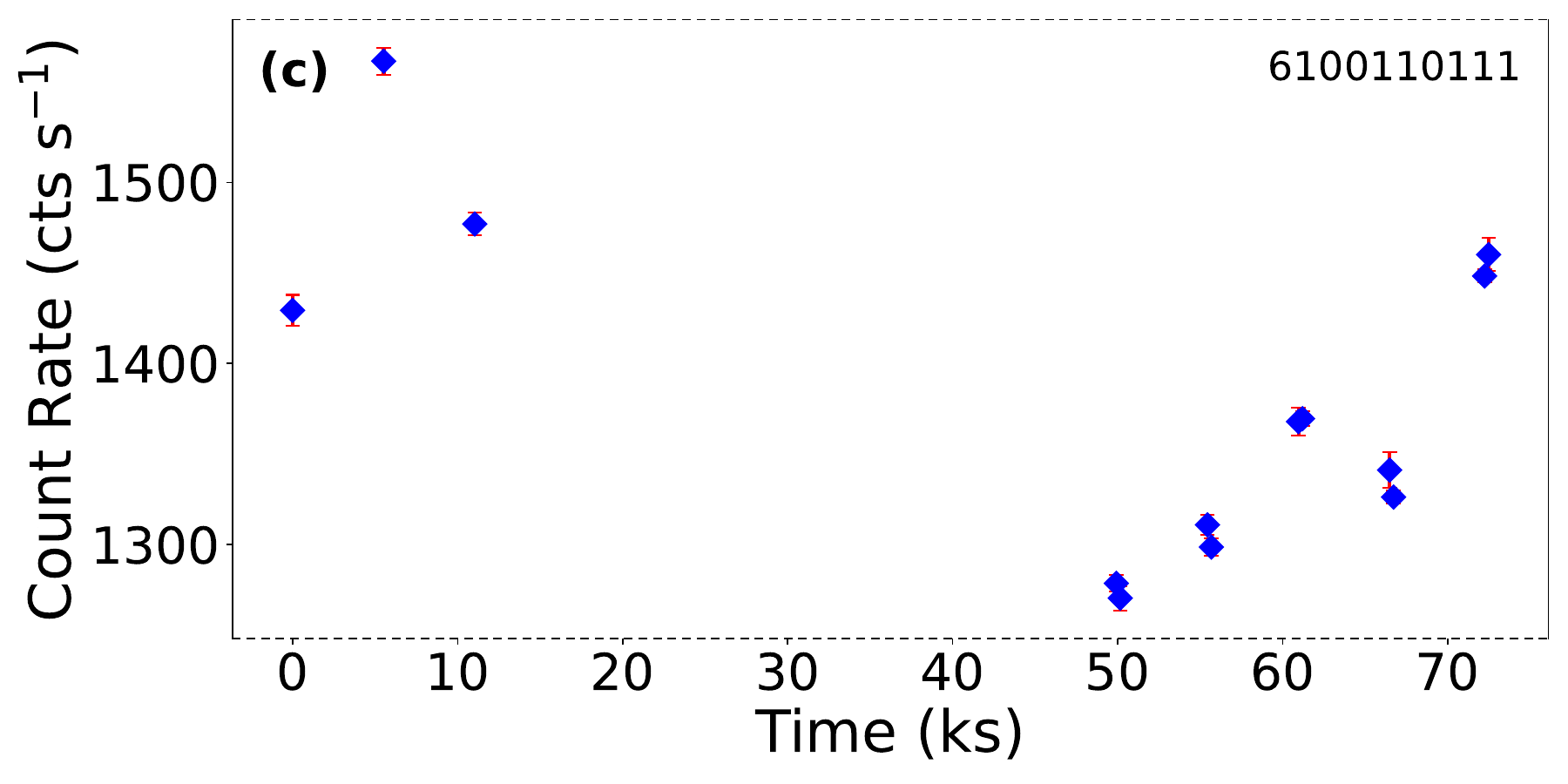}

\caption{\textit{NICER} observations of Mrk~421: (a) long-term X-ray light curve (2022–2024; each point is the mean count rate per observation, in chronological order as in Table~1); (b, c) light curves of two  individual \textit{NICER} observations.}
\label{fig:mean_count_rate}
\end{figure*}

For timing analysis, light curves were extracted from the barycenter-corrected event files using the \texttt{nicerl3-lc} task, with a bin size of 60~s.


\section{Analysis and Results}\label{sec:analysis}
We carried out detailed timing and spectral analyses of Mrk~421 using \textit{NICER} observations, aimed at probing its temporal variability and spectral evolution. A total of 45 \textit{NICER} observations with exposure times greater than 1000~s were selected to ensure data quality and consistency. The study is organized into two sections: (i) timing analysis, to characterize fluctuations over different time intervals, and (ii) spectral analysis, aimed at understanding the emission mechanisms and physical conditions driving the observed spectra.

\subsection{Timing analysis}

To investigate the temporal characteristics of the source, we extracted \textit{NICER} light curves in the 0.4--10~keV energy range as detailed in Section~\ref{sec:data}. A summary of all the \textit{NICER}     observations used in our study is presented in Table~\ref{tab:obs-details}, which lists the observation ID, exposure, date, mean count rate with error, and fractional variability. Figure~\ref{fig:mean_count_rate}(a) shows the long-term X-ray light curve of Mrk~421 in chronological order, based on two years of \textit{NICER} observations collected between 2022 and 2024. The mean count rate demonstrates significant variability, indicative of strong flux change over time, with flux levels increasing by approximately a factor of 28.~The substantial variability in flux reflects intense and temporally variable emission processes, likely governed by shifts in acceleration efficiency, magnetic-field configuration, or Doppler-boosting effects. The individual light curves in Figures~\ref{fig:mean_count_rate}(b) and (c) demonstrate significant intraday variability, with flux changes occurring on timescales of a few hours.
\subsubsection{RMS–Flux correlation}
\begin{figure}[!htbp]
  \centering
  \hspace*{-0.75cm}
  \includegraphics[width=\linewidth]{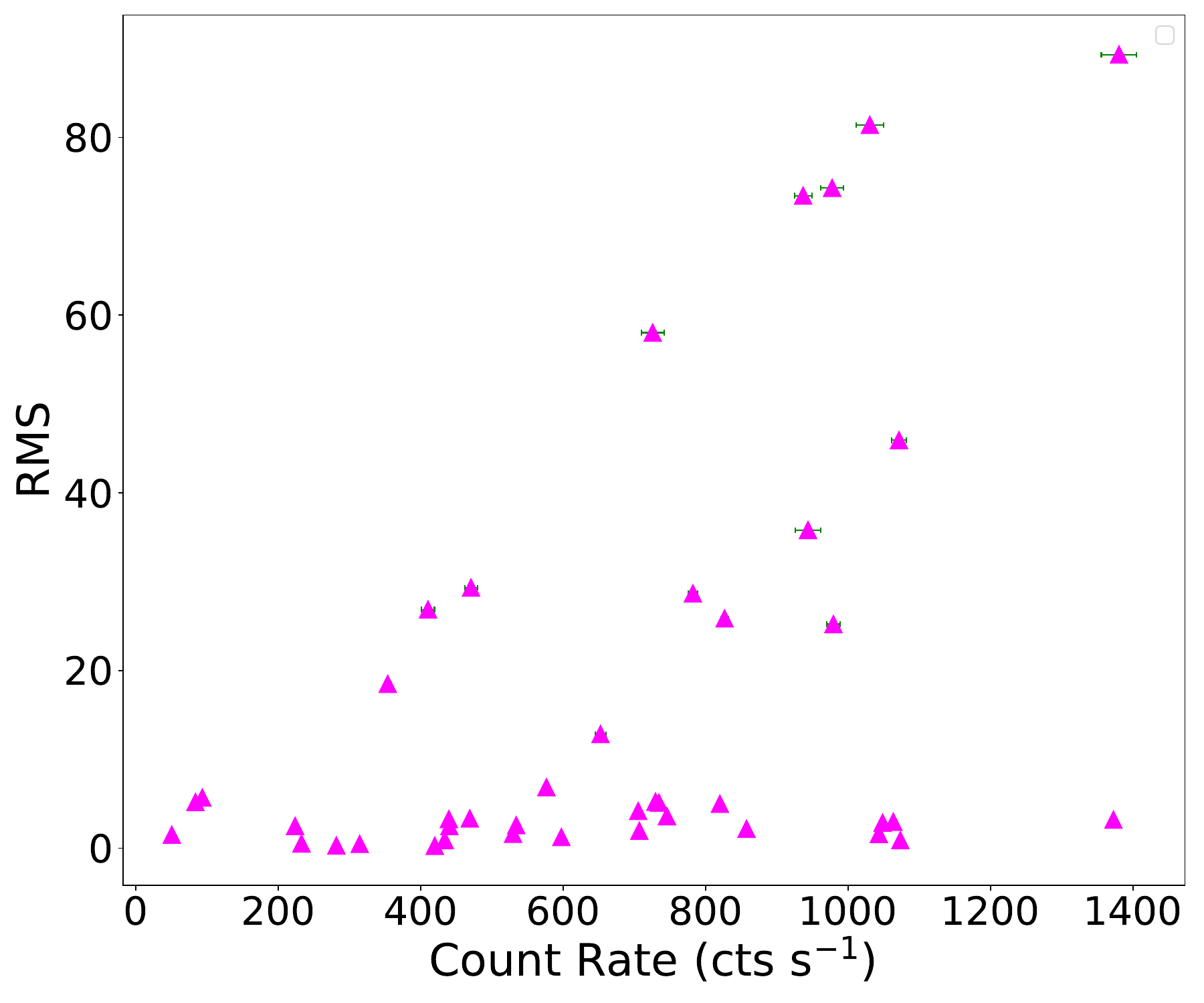}
  \caption{Root mean square (RMS) variability as a function of mean count rate for Mrk~421.}
  \label{fig:fig2}
\end{figure}

Investigating the relationship between root mean square (RMS) variability and mean count rate is a powerful tool for understanding the physical mechanisms driving high-energy astrophysical phenomena.~Identified initially in the X-ray variability of AGN and X-ray binaries, this relation suggests that short-term flux fluctuations scale in proportion to longer-term brightness variations, consistent with the behavior expected from accretion-driven processes \citep{bhatta2025probing, uttley2001flux, Uttley2005}. Excess variance quantifies the intrinsic variability of a source by subtracting the variance due to measurement errors from the overall variance observed in the light curve \citep{Nandra1997,Edelson2002}. The RMS amplitude of intrinsic variability is obtained by taking the square root of the excess variance, thus providing a measure of the absolute variability in the same units as the observed flux. For a set of $N$ observed flux values $X_i$, each with a corresponding uncertainty $\sigma_{\mathrm{err}}$, and a sample variance $S^2$, the excess variance is determined as follows:
\begin{equation} \sigma^2_{\mathrm{XS}} = S^2 - \overline{\sigma^2_{\mathrm{err}}}, \end{equation} where $\overline{\sigma^2_{\mathrm{err}}}$ represents the mean of the squared measurement uncertainties. The linear RMS–flux relationship was first identified in studies of variability in X-ray binaries containing accretion discs \citep{uttley2001flux, gleissner2004long, heil2012ubiquity}.~This relationship was subsequently reported in Seyfert galaxies \citep{vaughan2003characterizing, uttley2001flux}. For each observation, the mean count rate and the corresponding excess variance are used to quantify intrinsic variability.
More recently, similar correlations have emerged in blazar studies in the optical, X-ray, and gamma ray bands \citep{bhattacharyya2020blazar, bhatta2020nature, bhatta2021characterizing, kundu2022rms, edelson2013kepler, wang2023comprehensive}. The relationship between RMS and the count rate of Mrk~421 is illustrated in Figure~\ref{fig:fig2}, which does not reveal a clear RMS–flux relationship in the \textit{NICER} data of Mrk~421.
This discrepancy may arise from limited sampling and irregular observational cadence, which can bias the sampled flux range, as well as from the high sensitivity of \textit{NICER} in resolving both faint and bright states. Similar complexities in long-term X-ray variability have also been reported in other blazars \citep[see, e.g.,][]{bhatta2025probing,dinesh2023constraining}.
\subsubsection{Fractional variability}
To quantify both intraday and long-term variability of the source, we compute fractional variability using the following relation:
\begin{equation}
F_{\mathrm{var}} = \sqrt{\frac{S^2 - \bar{\sigma}_{\mathrm{err}}^2}{\bar{X}^2}}
\end{equation}
\citep{vaughan2003characterizing,edelson1990broad}. 
Here, \( S^2 \) is the total variance of the light curve, $\bar{\sigma}_{\mathrm{err}}^2$ is the mean square of the measurement uncertainties, and $\bar{X}$ is the mean flux. 
The associated error in fractional variability is obtained using the following equation,
\begin{equation}
\sigma_{F_{\mathrm{var}}} = \sqrt{
\left( \frac{1}{\sqrt{2N}} \cdot \frac{\bar{\sigma}_{\mathrm{err}}^2}{F_{\mathrm{var}} \bar{X}^2} \right)^2
+ \left( \sqrt{\frac{\bar{\sigma}_{\mathrm{err}}^2}{N}} \cdot \frac{1}{\bar{X}^2} \right)^2 }
\end{equation}
\citep{bhatta2018microvariability, Aleksic2015}.

For the 45 X-ray light curves, fractional variability was derived using the above relation, and the results are summarized in Column 6 of Table~\ref{tab:obs-details}. The source exhibits a total fractional variability of roughly 48\% across all observations. The table shows that the highest fractional variability, $F_{\mathrm{var}} = 7.99 \pm 0.09\%$, was recorded on 2023-05-24, indicating a period of strong X-ray activity. In contrast, the lowest variability ($F_{\mathrm{var}} = 0.11 \pm 0.29\%$) occurred on 2024-03-19, corresponding to a comparatively quiescent state.

\subsubsection{ Flux Distribution }
The flux distribution characterizes how frequently a variable source occupies different flux states across several observations. The form of the probability distribution function (PDF) that best describes these distributions reflects the underlying physical processes. In active galactic nuclei (AGN) in particular, additive processes generally produce Gaussian (normal) distributions, whereas multiplicative processes tend to yield log-normal profiles. Extensive studies have characterized these distributions in blazars across multiple energy bands and time scales \citep{bhattacharyya2020blazar, bhatta2021characterizing, mohorian2022x, dinesh2023constraining, bhatta2020nature}. The distribution is clearly multi-peaked and cannot be described by a single functional form such as a normal or log-normal distribution.  To quantitatively characterize the observed distribution profile of the X-ray count rates ($x$), we model the distribution using a superposition of three Gaussian components of the form

\begin{equation}
P(x)=\sum_{i=1}^{3} A_i
\exp \left[
-\frac{(x-\mu_i)^2}{2\sigma_i^2}
\right]
\end{equation}
where $A_i$, $\mu_i$, and $\sigma_i$ represent the amplitude, centroid, and width of the $i$-th Gaussian component, respectively. A combination of three Gaussian components provides a good representation of the data, with peaks at $\sim423$, $\sim822$, and $\sim1065$ counts s$^{-1}$. These correspond to low, intermediate, and high flux states, respectively.
\begin{figure}[H]
    \centering
    \includegraphics[width=\linewidth]{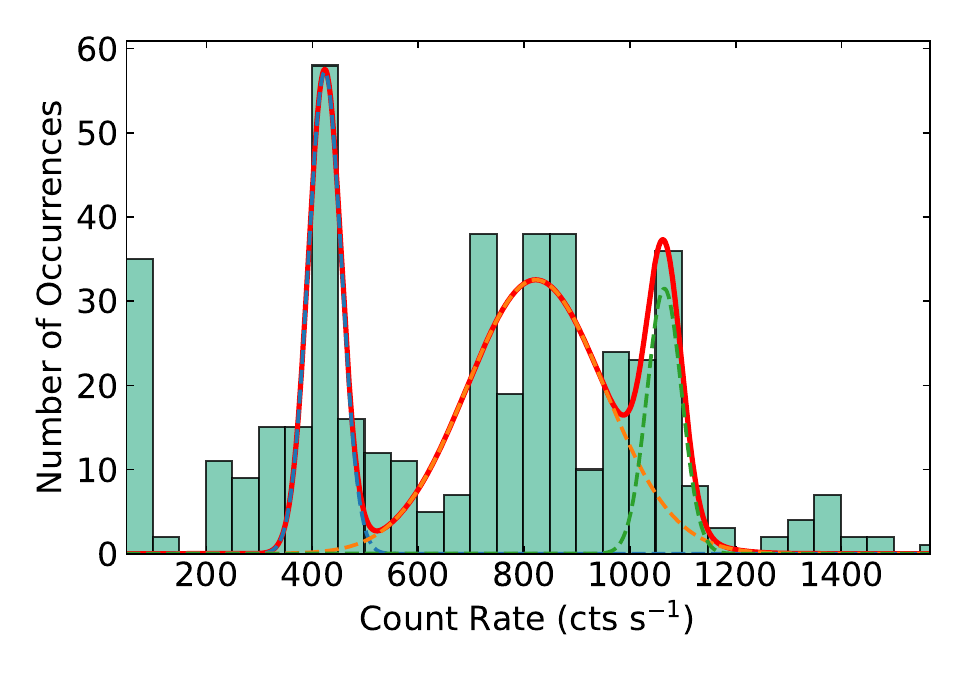}
    \caption{Histogram depicting the distribution of X-ray flux values for Mrk~421, observed with \textit{NICER} between 2022 and 2024. The multimodal profile is fitted with three Gaussian components, representing distinct low-, intermediate-, and high-flux states.}
    \label{fig:flux_histogram}
\end{figure}

The presence of multiple distinct peaks indicates that the variability of Mrk 421 is not driven by a single stochastic process. Instead, this multimodal behavior suggests the existence of multiple emission states, likely associated with different physical conditions within the jet. These states may correspond to variations in particle acceleration efficiency, magnetic field strength, or contributions from multiple emitting regions.

\subsection{Spectral analysis}
We perform spectral analysis of \textit{NICER} observations of Mrk~421 using both model-independent and model-dependent techniques.~The hardness ratio was first calculated to empirically examine the spectral variability. This was followed by a detailed spectral fitting procedure using the XSPEC software package \citep{arnaud1996}, allowing the derivation of physical parameters and offering insight into the nature of the source's X-ray emission. We fixed the Galactic hydrogen column density at 
$N_{\mathrm{H}} = 0.0134 \times 10^{22}\,\mathrm{cm}^{-2}$ 
to account for line-of-sight absorption using the HEASARC tool $N_{\mathrm{H}}$ \footnote{\url{https://heasarc.gsfc.nasa.gov/cgi-bin/Tools/w3nh/w3nh.pl}}.

\subsubsection{Hardness Ratio (HR)}
To explore the interdependence between soft and hard X-ray emissions in Mrk~421 and to investigate spectral variability, we extracted light curves in two distinct energy bands: the soft band (LE; 0.4--1~keV)  and the hard band (HE; 1--10~keV). The hardness ratio (HR), a model-independent diagnostic of spectral shape, is widely employed to quantify spectral changes resulting from variations in the physical conditions within the relativistic jet or due to accretion-related processes onto the central supermassive black hole. In addition, HR is often used to classify blazars based on spectral hardness and to monitor their temporal evolution.
In this study, the hardness ratio is defined as follows:

\setlength{\abovedisplayskip}{6pt}
\setlength{\belowdisplayskip}{6pt}
\begin{equation}
\label{hard}
    HR = \frac{H}{S},
\end{equation}
where $H$ and $S$ are the count rates in the hard and soft energy bands, respectively. The associated uncertainty in the hardness ratio, $\sigma_{\rm HR}$, is calculated using:

\setlength{\abovedisplayskip}{6pt}
\setlength{\belowdisplayskip}{6pt}
\begin{equation}
\sigma_{\rm HR}
= \frac{H}{S}\,
\sqrt{
\left( \frac{\sigma_H}{H} \right)^{2}
+
\left( \frac{\sigma_S}{S} \right)^{2}
},
\end{equation}
where $\sigma_H$ and $\sigma_S$ are the uncertainties in $H$ and $S$, respectively.

\begin{figure}[]
    \centering
    \includegraphics[width=\linewidth]{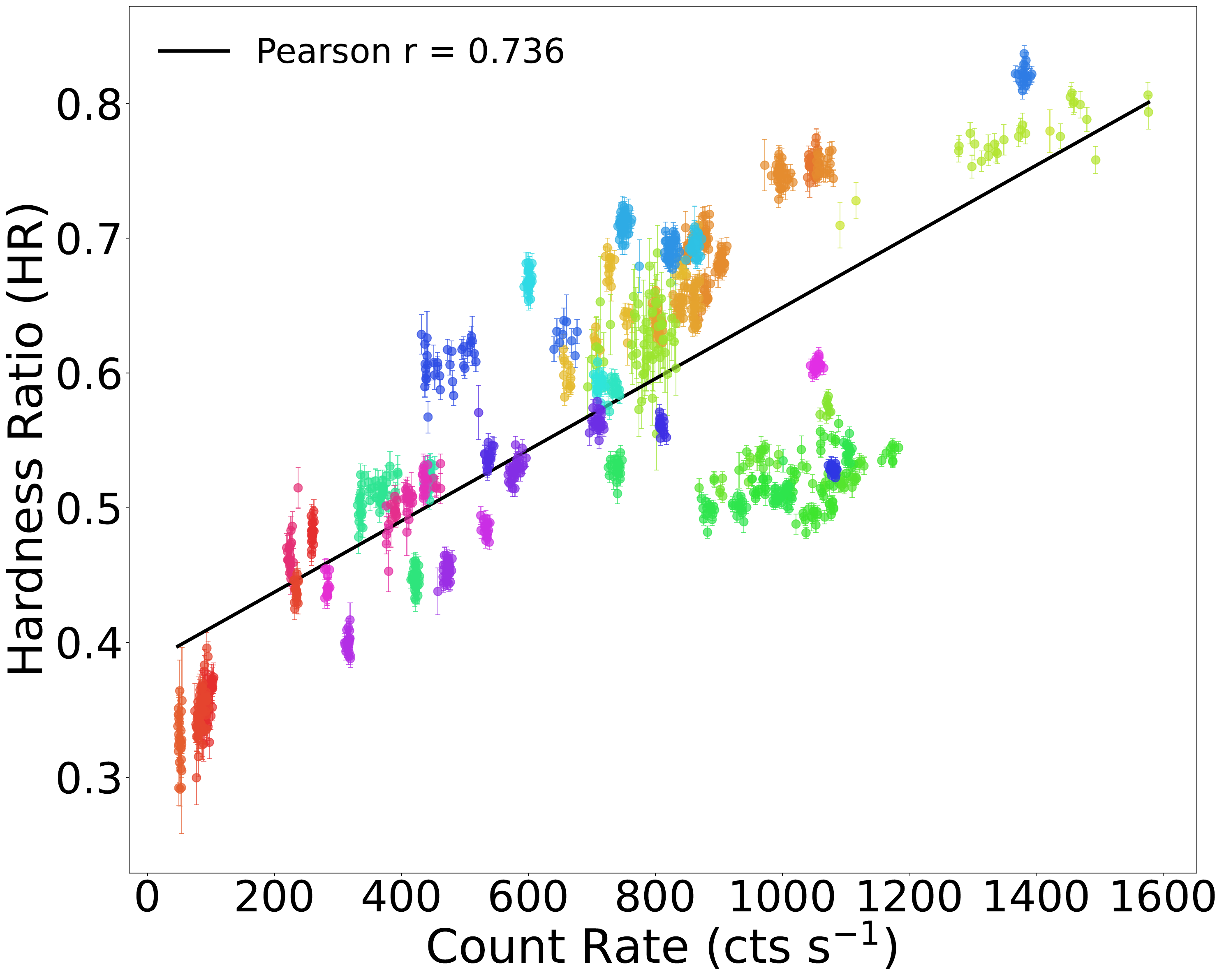}
    \caption{Distribution of the hardness ratio (HR) as a function of count rate (in counts per second) for Mrk~421. Each color represents a different observation ID, and the solid black line shows the best-fit linear correlation (Pearson $r = 0.73$). The green points deviate significantly from the general trend and are considered outliers.}
    \label{fig:HR}
\end{figure}

Variations in HR directly reflect spectral changes over time. We analyzed the HR versus total count rate across the full 0.4--10~keV energy range. The resulting distribution is depicted in Figure~\ref{fig:HR}, which illustrates the significant correlation (Pearson $r = 0.736$) between the HR and the X-ray count rate in the full band. The best-fit line between the count rate and HR, shown by a solid black line exhibits a clear linear trend, the hardness for bright states getting as high as 0.8.  The increasing HR with the count rate for most of the sources indicates that high-energy emission becomes increasingly prominent relative to the soft band.
The linear trend widely termed as \textit{harder-when-brighter} and commonly observed in several blazars is likely driven by a more efficient particle acceleration mechanism at higher flux states. \citep[see][and the references therein]
{bhatta2018hard}.

This behavior is consistent with flare-like activity, where high-energy electrons are injected more efficiently, hardening the X-ray spectrum. We note that a subset of data deviates from this general relation, with the majority of these outliers appearing in green. These points likely represent intervals of unusual spectral behavior, possibly linked to transient flaring episodes or atypical emission states of Mrk 421.

\begin{figure}[H]
    \centering
    \includegraphics[width=\linewidth]{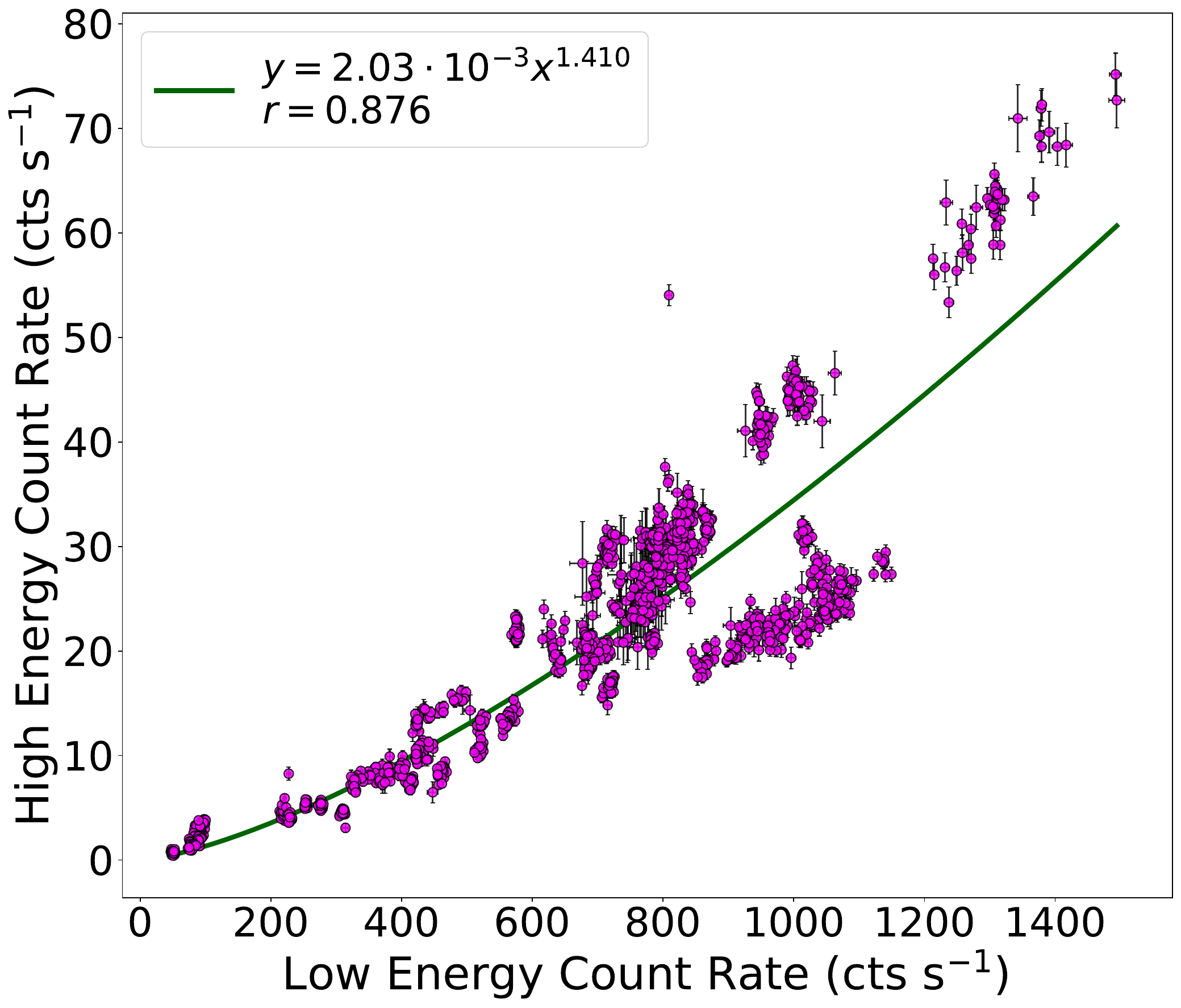}
\caption{Correlation between high energy and low energy count rates of Mrk~421 obtained from \textit{NICER} observations. The solid line represents the best-fitting power-law model.}
    \label{fig:HE-LE}
\end{figure}

Moreover, we also observed a strong correlation (Pearson $r = 0.91$) between the HE light curve versus the LE light curve, implying that both bands originate from the same emission region. Figure~\ref{fig:HE-LE} shows the distribution of observations in the two bands, with a power-law fit $\sim$1.17  representing the best-fit trend.

\begin{figure}[H]
    \centering
    \includegraphics[width=0.9\columnwidth]{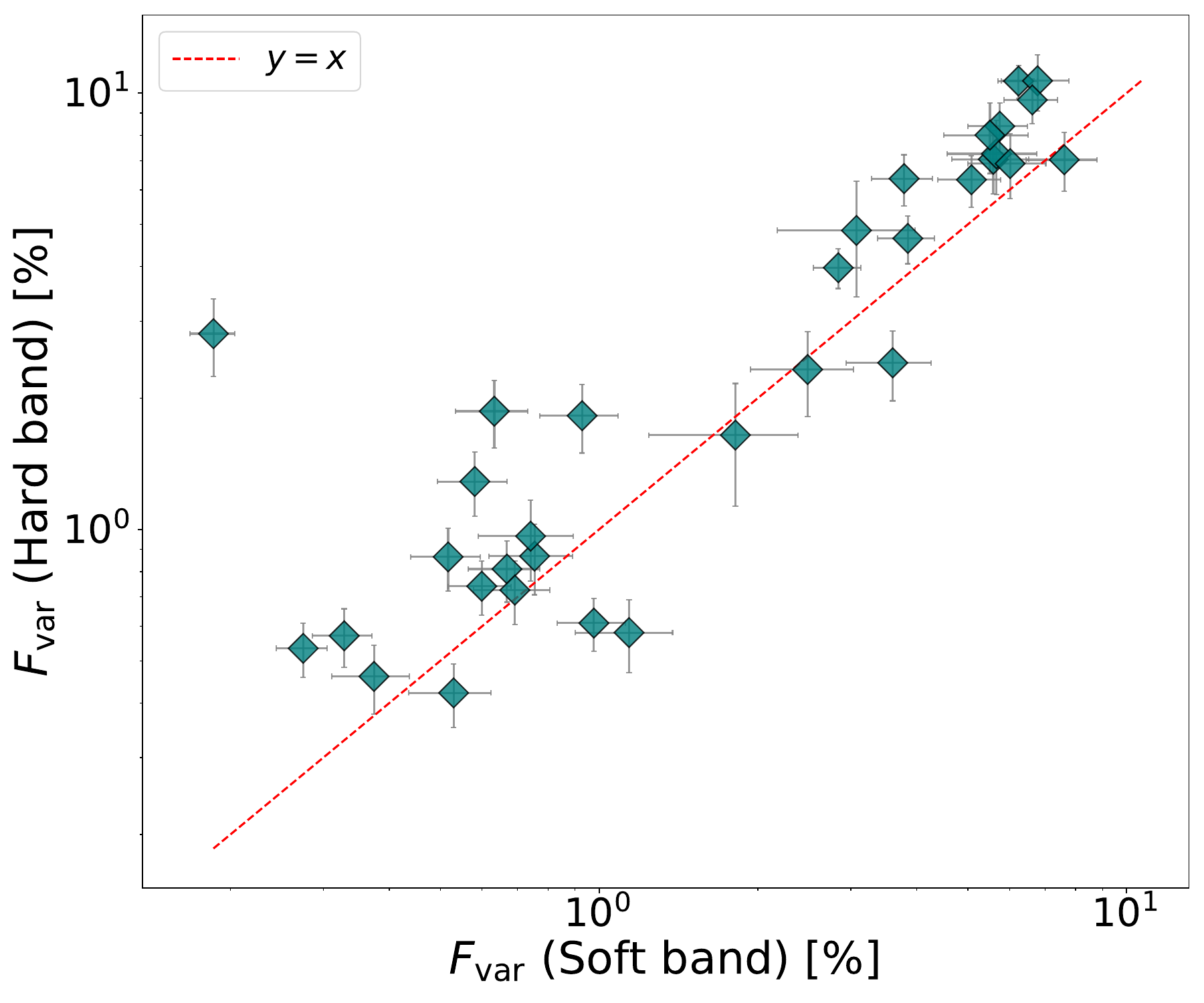}
    \caption{Relation between high-energy and low-energy fractional variability for Mrk~421. The red line represents $y=x$ for comparison.}
    \label{fig:Fvarcomp}
\end{figure}
\vspace{1em}

Furthermore, Figure~\ref{fig:Fvarcomp} shows the distribution of $F_{\mathrm{var}}$ between both LE and HE X-ray bands. The red line represents $y=x$ for comparison. It can be seen that the fractional variability in the HE band is generally higher than that in the LE band, consistent with the synchrotron nature of the X-ray emission.

\subsubsection{Spectral Modeling }
\label{spec_fit}
To investigate the non-thermal emission properties of Mrk 421, we analyzed 45 \textit{NICER} observations listed in Table \ref{tab:obs-details} using the PyXspec spectral fitting package in accordance with the standard X-ray spectral analysis procedures \citep{Gendreau2016,arnaud1996}.
~To minimize instrumental uncertainties and optimize the accuracy of calibration, the analysis was performed in the 0.4–10 keV band. For each observation, we extracted the source spectrum along with the RMF and ARF files.~To account for the effects of galactic absorption, we incorporated the T\"ubingen--Boulder ISM absorption model
 (\textit{tbabs}) into each spectral fit \citep{arnaud1996, Wilms2000}.

In order to assess the spectral behavior of Mrk 421, we applied three spectral models to characterize the X-ray emission from Mrk 421: a power-law (PL), a log-parabola (LP) and a broken power-law (BPL).
\vspace{6pt}

(i) Power-law (PL): 
A frequently employed spectral model in high-energy astrophysics is the PL model, which characterizes the photon count distribution across energy as a simple PL function. The functional form is given by
\begin{equation}
\frac{dN}{dE} = K \cdot E^{-\Gamma},
\tag{7}
\end{equation}
where \( K \) denotes the normalization constant and \( \Gamma \) is the photon index that governs the slope of the spectrum.

\vspace{6pt}
ii) Broken Power-Law (BPL): The BPL model describes a spectrum characterized by two PL segments exhibiting a change in slope at a characteristic energy, referred to as the break energy, and is defined as:
\begin{equation}
\frac{dN}{dE}=
\begin{cases}
K\,E^{-\Gamma_1}, & E \le E_{\rm break},\\
K\,E_{\rm break}^{\,\Gamma_2 - \Gamma_1}\,E^{-\Gamma_2}, & E > E_{\rm break},
\end{cases}
\tag{8}
\end{equation}
where \( K \) is the normalization constant, \( \Gamma_1 \) and \( \Gamma_2 \) are the photon indices below and above \( E_b \), respectively. A BPL spectrum signifies a population of energetic particles, in which high-energy relativistic electrons undergo rapid radiative cooling, resulting in a spectral break.
\vspace{6pt}

(iii) Log-Parabola (LP): The LP model is widely used in the spectral analysis of blazars, because of its ability to capture curved spectral shapes. The photon distribution is expressed as:  
\begin{equation}
\frac{dN}{dE} = K \left( \frac{E}{E_1} \right)^{ -\alpha - \beta \log \left( \frac{E}{E_1} \right) }
\tag{9}\label{eq:logpar},
\end{equation}
where $K$ is the normalization constant, $E_1$ is the pivot (or reference) energy fixed at 1 keV during spectral fitting, $\alpha$ is the photon index at $E_1$, and $\beta$ is the curvature parameter that quantifies the degree of spectral curvature.~The spectral curvature described by the LP is commonly interpreted as the outcome of stochastic acceleration, where particles have a reduced probability of gaining energy at higher energies \citep{massaro2004log,massaro2006}.
Furthermore, the peak energy $E_{\rm p}$, at which the flux reaches its maximum, is given by, 
\begin{equation}
E_{\rm p} = E_1 \times 10^{\frac{2-\alpha}{2\beta}}
\tag{10}
\end{equation}
The LP model defined in Eq.~\ref{eq:logpar} is widely used to describe curved X-ray spectra, in the emission of AGN particularly in cases where no sharp high-energy cut-off, such as that produced by an exponential form, is observed.~By introducing only one additional parameter \noindent relative to a PL representation, the LP model is particularly useful for estimating quantities of physical relevance to the emission region, such that it is possible to define an energy-dependent photon index, $\Gamma(E)$, obtained as the logarithmic derivative of Equation~\ref{eq:logpar}:

\begin{figure*}[ht!]
\gridline{
  \fig{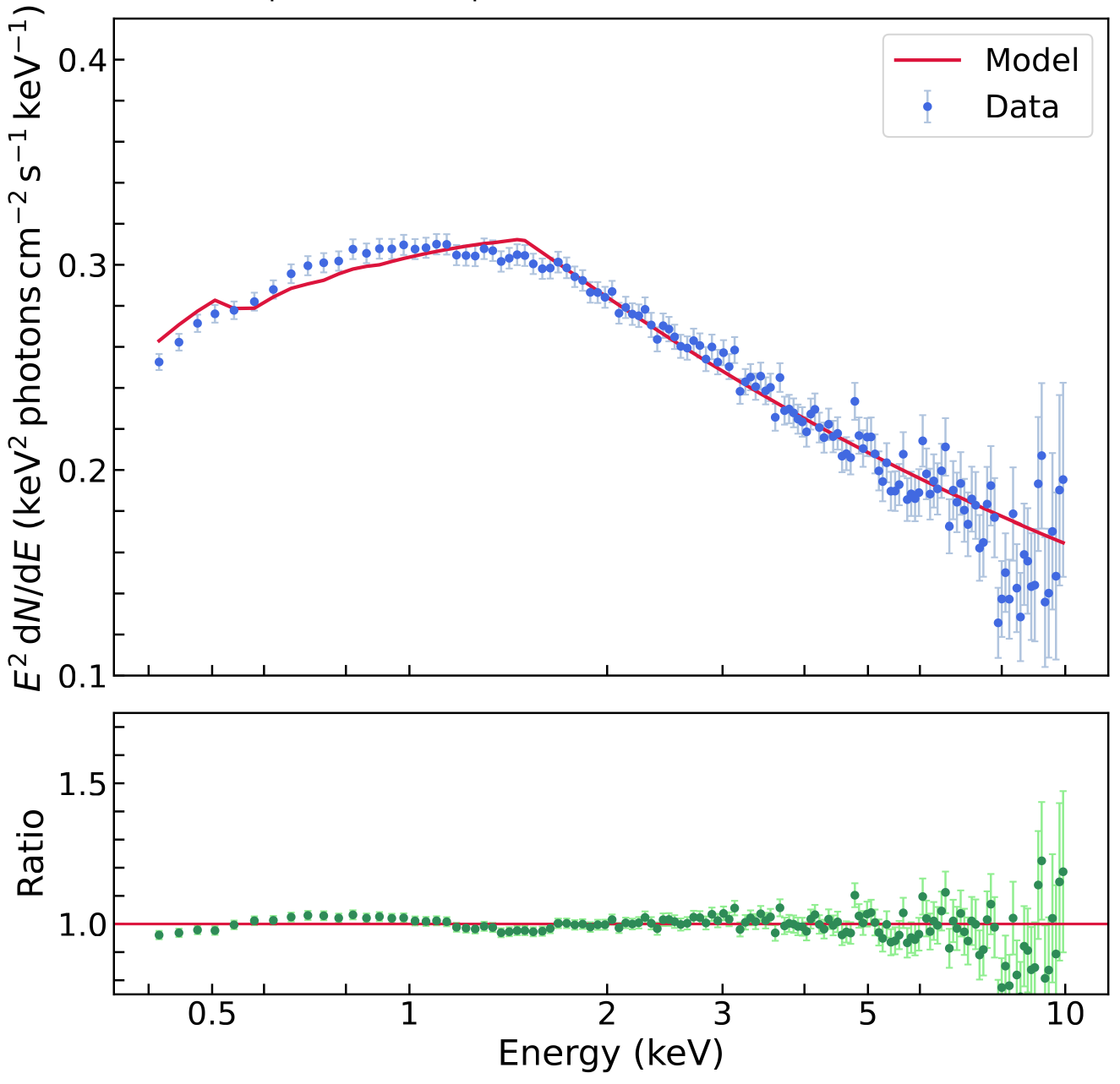}{0.45\textwidth}{(a) 6704016901 – Broken Power-Law}
  \fig{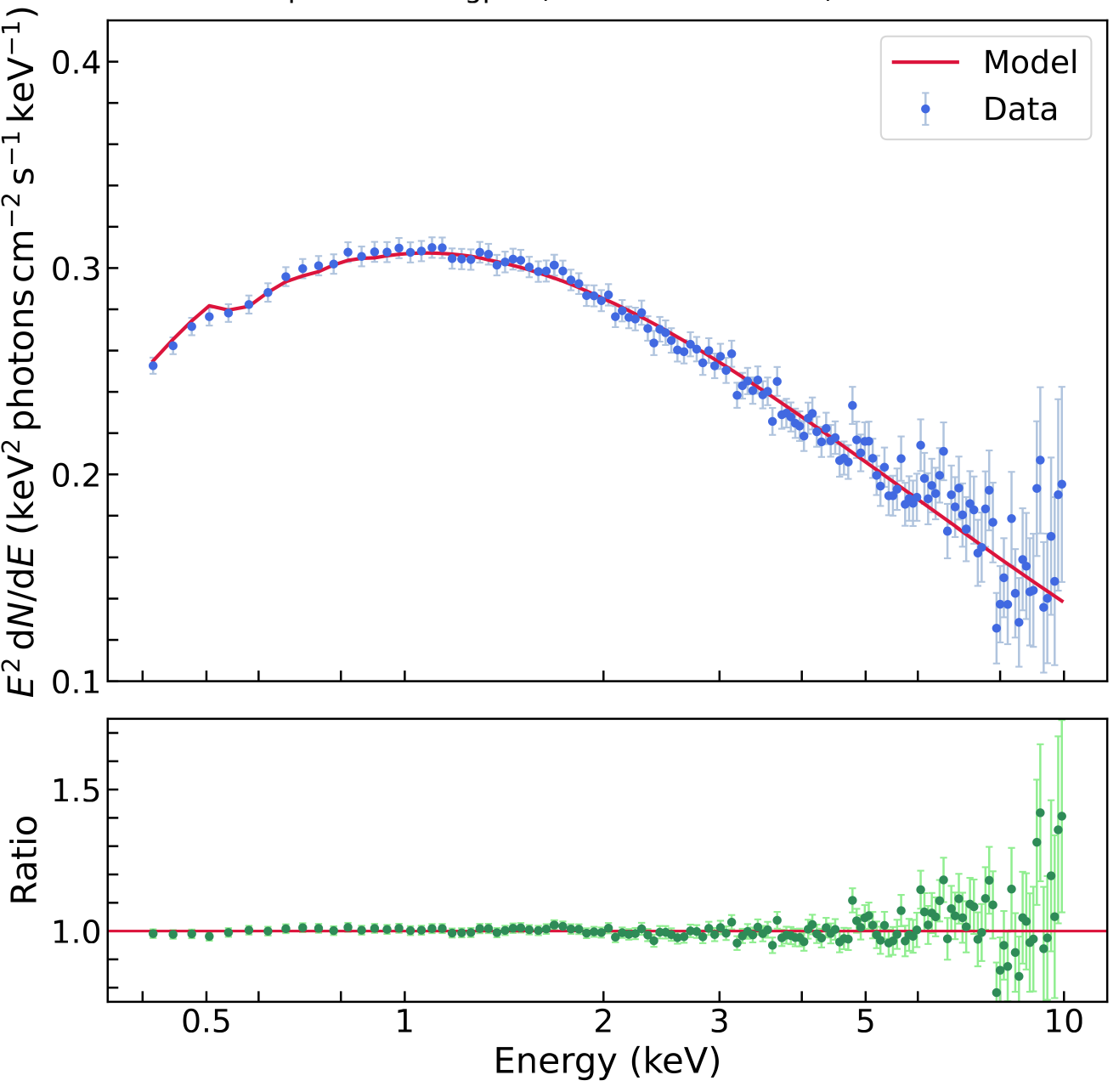}{0.45\textwidth}{(b) 6704016901 – Log-Parabola}
}
\gridline{
  \fig{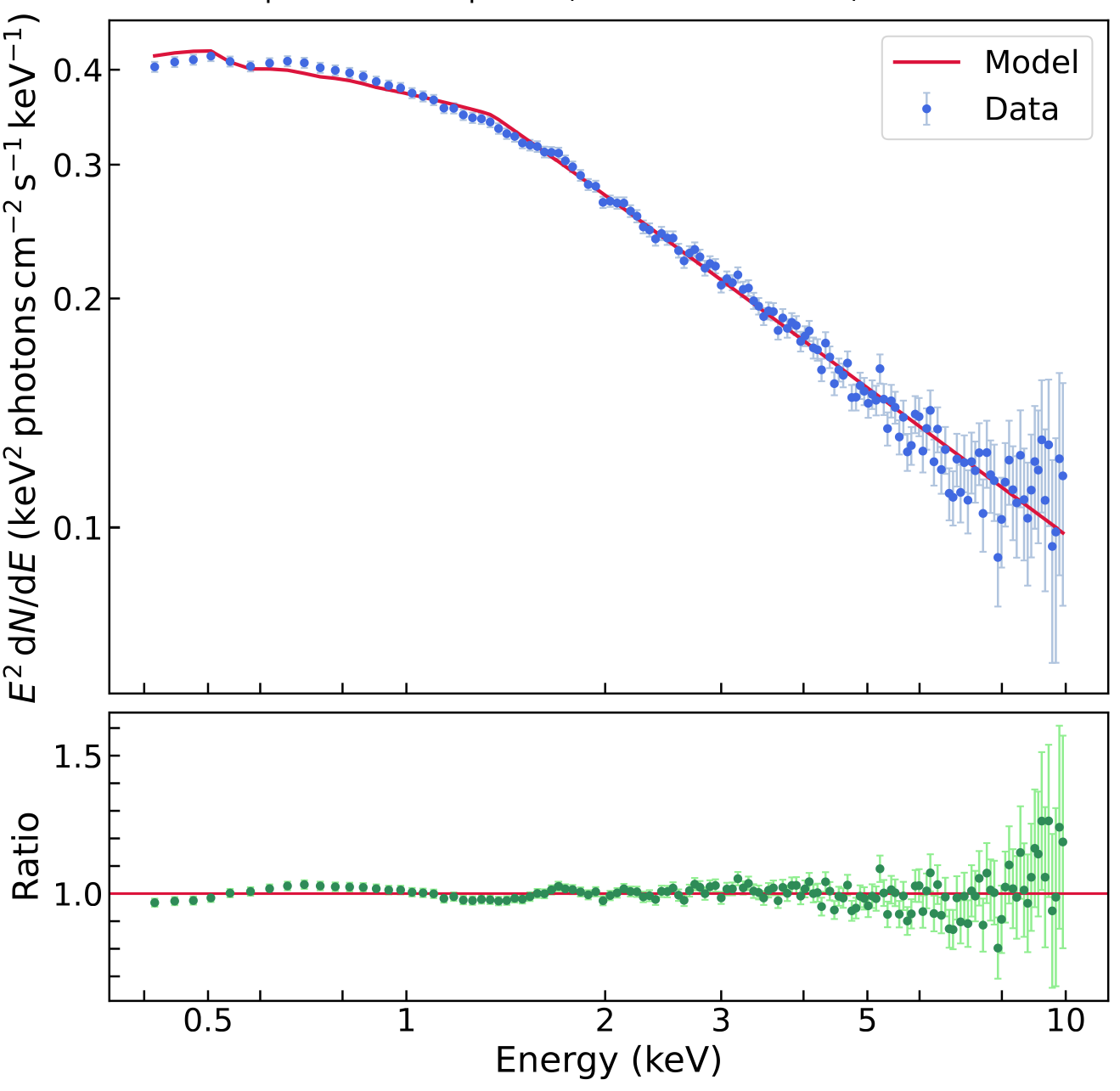}{0.45\textwidth}{(c) 6100110135 – Broken Power-Law}
  \fig{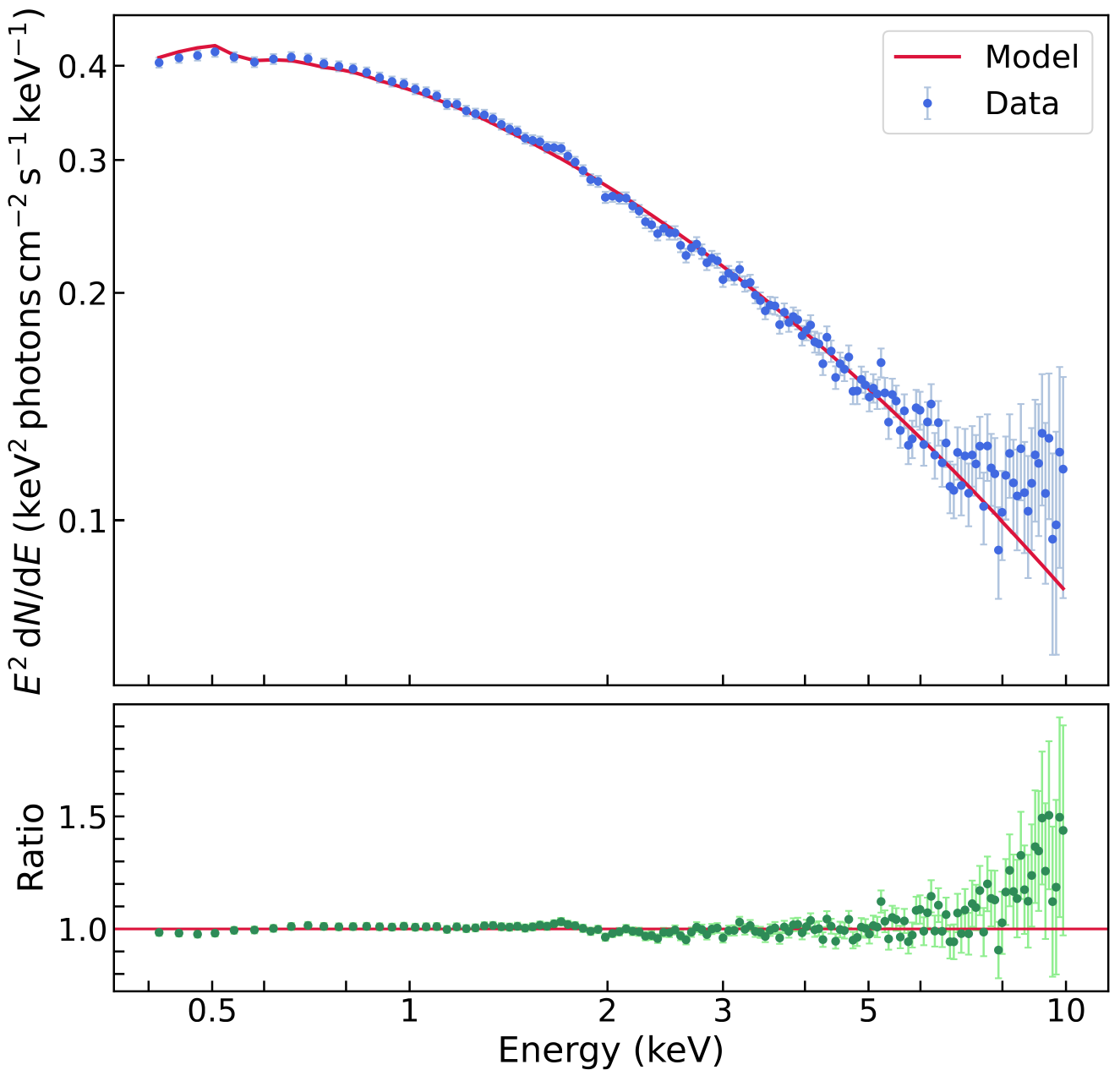}{0.45\textwidth}{(d) 6100110135 – Log-Parabola}
}
\caption{Comparison of spectral fits using broken power-law and log-parabola models 
for two \textit{NICER} observations of Mrk~421.}
\label{fig:figspec}
\end{figure*}

\begin{equation}
\Gamma(E) = \alpha + 2\beta \,\log_{10}\!\left(\frac{E}{E_{1}}\right).
\tag{11}\label{eq:logpara}
\end{equation}
A simple PL model was initially employed, but yielded relatively higher reduced chi-square values ($\chi^2_r$) for most observations. To improve the fit quality, we then employed two more complex models: the BPL and the LP. Both models significantly improved the $\chi^2_r$ values, indicating better agreement with the observed spectra compared to the PL model.

Furthermore, we used the F-test to compare alternative spectral models, following the standard approach described in \citet{Bevington1992}. This method compares the improvement in $\chi^{2}$ relative to the loss of degrees of freedom, and thus evaluates whether the more complex, nested model provides a genuinely better fit than the simpler one.
The F-test statistic is defined as

\begin{equation}
\tag{12}
F = \frac{(\chi^{2}_{1} - \chi^{2}_{2})/(d_{1} - d_{2})}{\chi^{2}_{2}/d_{2}},
\end{equation}
where $\chi^2_1$ and $d_1$ correspond to the chi-square and degrees of freedom of the simpler model, and $\chi^2_2$ and $d_2$ are those of the more complex, nested model. This statistic follows an $F$-distribution with $(d_1 - d_2, d_2)$ degrees of freedom, and the associated $p$-value gives the probability that the observed improvement arises by chance.

Table~\ref{tab:spectral-params} summarizes the results of the spectral fits, listing the observation ID and the applied \texttt{XSPEC} model in the first two columns. The third, fourth, and fifth columns contain the best-fit spectral parameters for each model, followed by the reduced chi-square (\( \chi_r^2 \)) values for each fit. The final column reports the F-test statistic together with the corresponding probability.

We applied the F-test, as available in \texttt{XSPEC}, to determine whether the improvement in fit of one model over the other was statistically significant. Of the total 45 \textit{NICER} X-ray spectra analyzed for Mrk 421, 42 observations were best described by the LP model, suggesting that the spectra frequently exhibit curvature. The result supports acceleration mechanisms in which the probability of energy gain decreases as the energy of the particle increases. The remaining three were better fitted using a broken power-law model based on the F-test, indicating occasional spectral breaks, possibly due to rapid cooling of high-energy electrons. The preference for LP and BPL models indicates that the X-ray emission of blazar jets is shaped by complex non-thermal processes \citep{massaro2004log}.
Representative fitted spectra for two observations are presented in Figure~\ref{fig:figspec}, which illustrates the best-fitting models applied to the data.
\subsubsection{Correlation studies}
\begin{figure*}[ht!]
\gridline{
  \fig{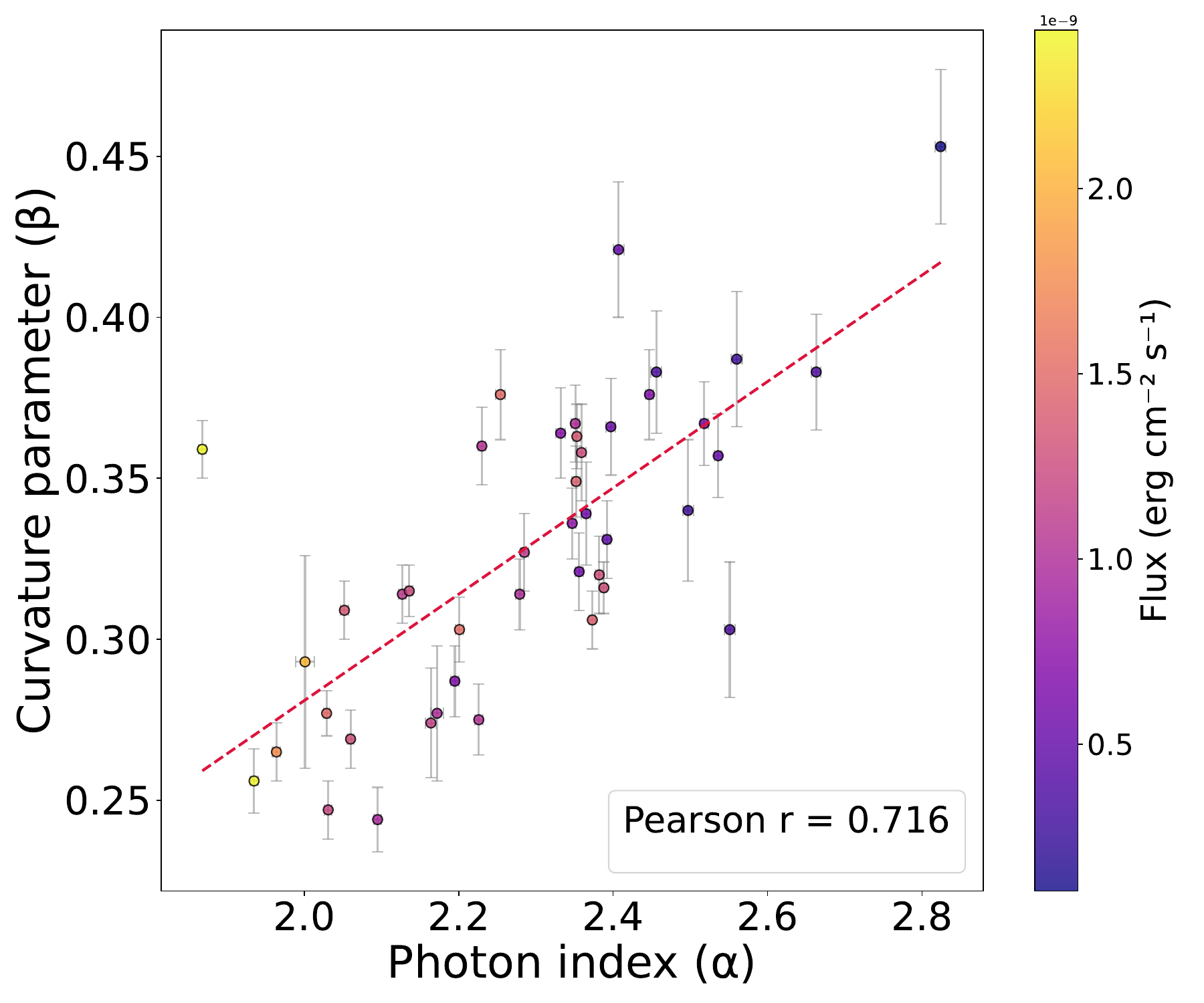}{0.42\textwidth}{(a)}
  \fig{Spectral_review/Ep_vs_flux.pdf}{0.42\textwidth}{(b)}
}
\gridline{
  \fig{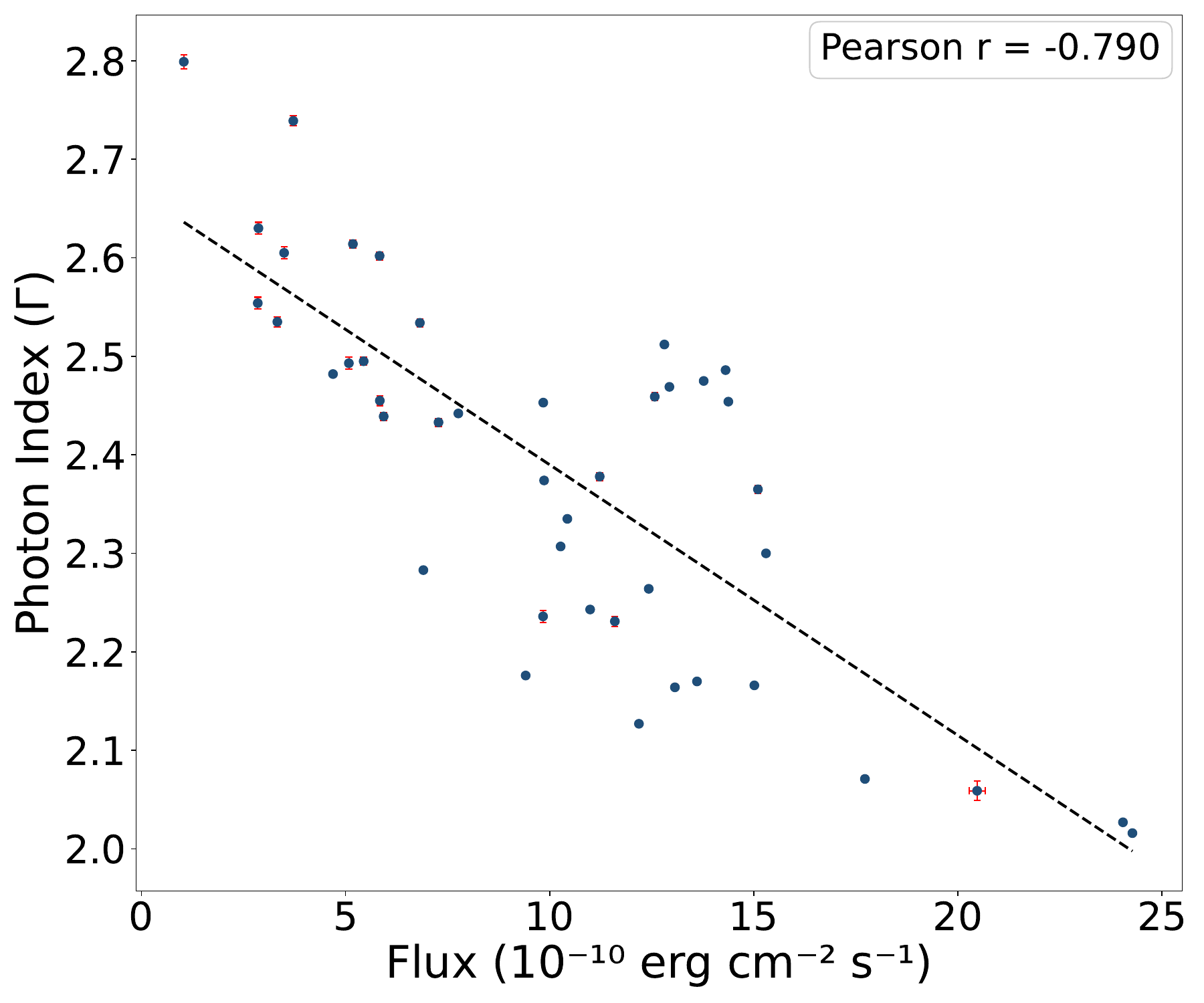}{0.42\textwidth}{(c)}
  \fig{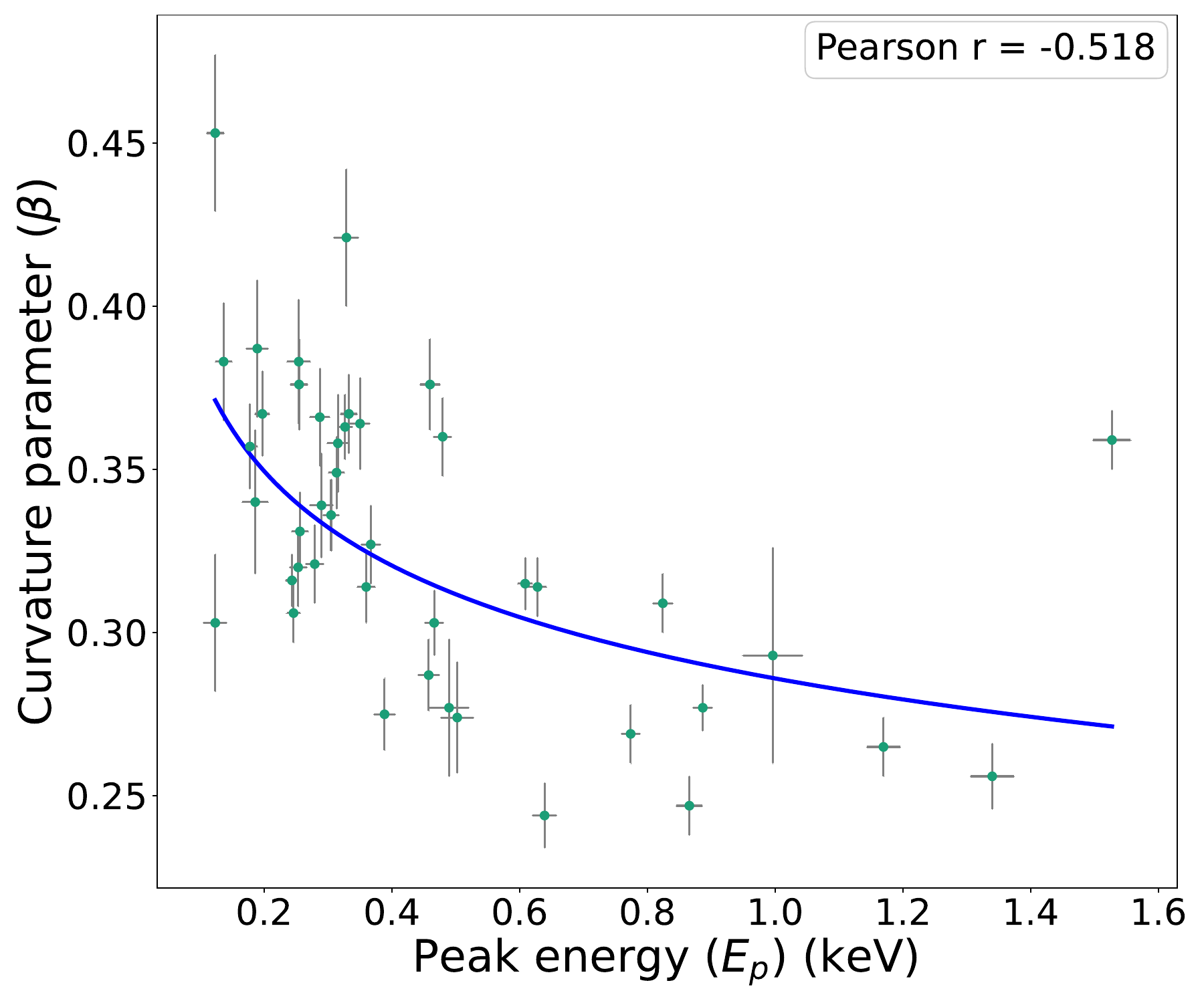}{0.42\textwidth}{(d)}
}
\gridline{
  \fig{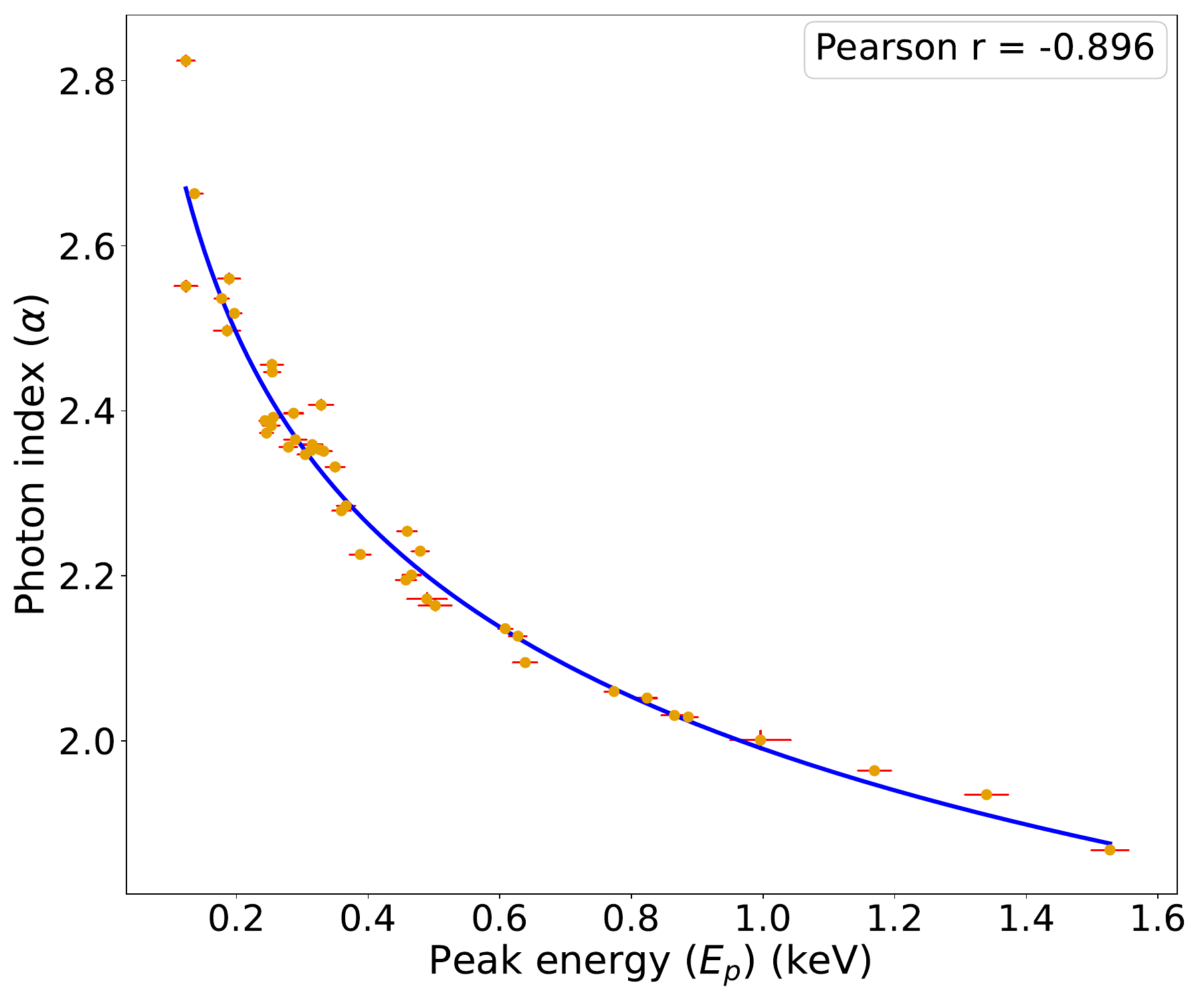}{0.42\textwidth}{(e)}
  \fig{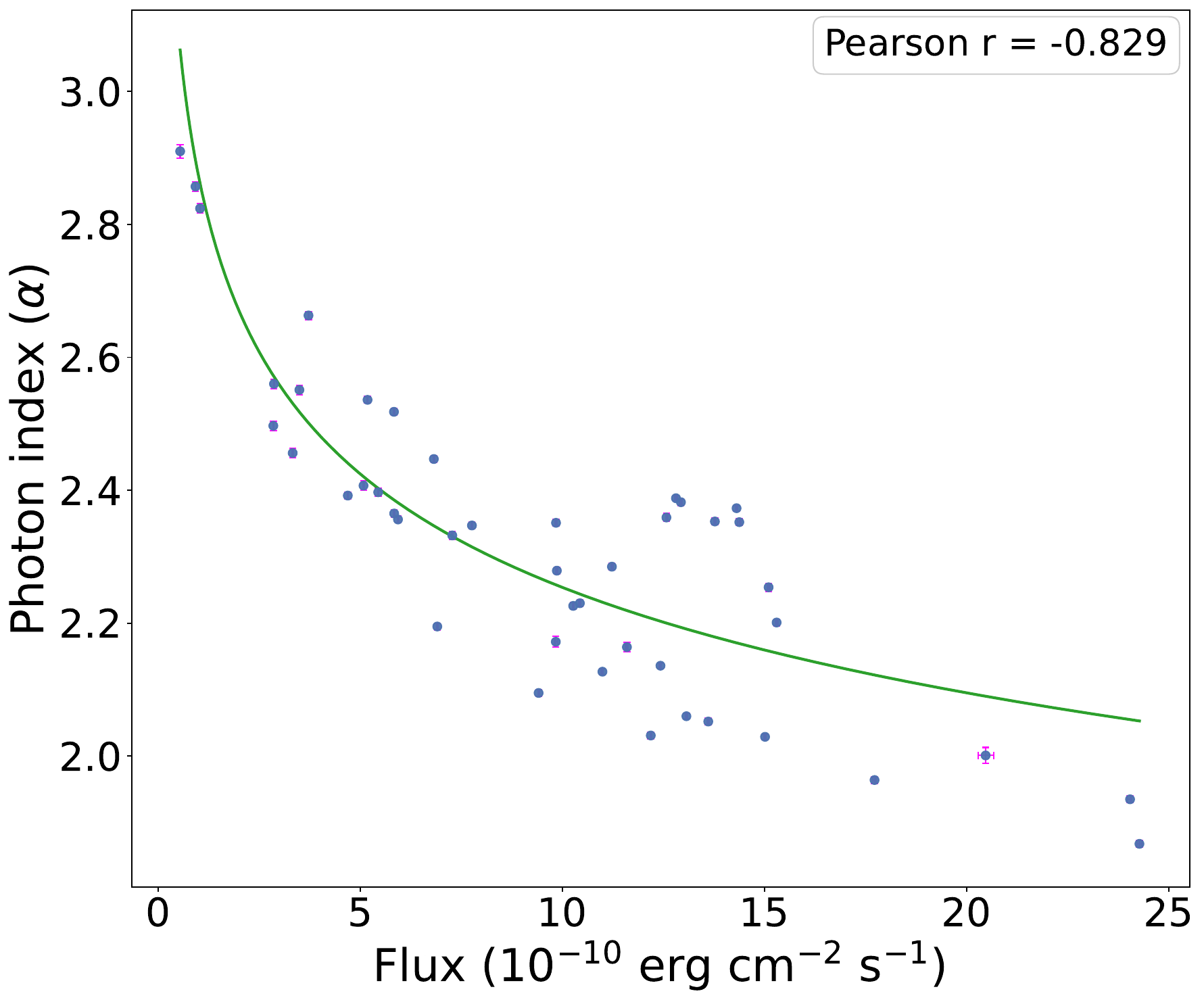}{0.42\textwidth}{(f)}
}
\caption{
Correlation plots showing the relations between spectral parameters and flux for Mrk~421:
(a) Log-parabolic photon index $\alpha$ versus curvature parameter $\beta$; the red dashed line shows the best-fit linear regression, and the symbols are color-graded according to flux.
(b) Synchrotron peak energy $E_{\mathrm{p}}$ versus flux, with the red dashed line indicating the best-fit trend,
(c) Power-law photon index  $\Gamma$ as a function of flux, with the green dashed line representing the best-fit relation,
(d) $E_{\mathrm{p}}$ versus $\beta$; the solid blue line shows the best-fit inverse power-law relation,
(e) $\alpha$ versus $E_{\rm p}$; the solid blue line represents the best-fit inverse power-law relation,
(f) $\alpha$ versus flux, with the blue line showing the best-fit inverse power-law relation.
These plots collectively illustrate the spectral variability of Mrk~421 and the evolution of spectral curvature with changing flux during the observed epochs.
}
\label{fig:corr}
\end{figure*}

To characterize the spectral behavior of Mrk 421 across a range of flux states, we examined the best-fit parameters listed in Table~\ref{tab:spectral-params} and investigated their correlations to identify consistent patterns of variability.
Figure~\ref{fig:corr}(a) shows the relationship between the photon index $\alpha$ and the curvature parameter $\beta$ with a significant positive correlation (Pearson \( r = 0.72 \)) which  is consistent with stochastic acceleration scenarios, where higher acceleration efficiency produces harder and less curved spectra, while reduced efficiency leads to softer and more curved emission \citep{tramacere2007signatures, massaro2004log, Tramacere2011stochastic}.~This supports the interpretation that the observed curvature originates  from the underlying particle acceleration processes in the jet.

Figure~\ref{fig:corr}(b) shows a strong positive correlation between the synchrotron peak energy ($E_p$)and the X-ray flux ($r = 0.82$), indicating that the synchrotron peak shifts to higher energies during brighter states. This implies that the emitting electrons attain higher characteristic energies. In Figure~\ref{fig:corr}(c), the photon index $\Gamma$ is plotted against the flux, revealing a pronounced anticorrelation ($r = -0.79$), indicating that the photon index decreases with increasing flux. This trend also supports \textit{harder-when-brighter}, which suggests spectral hardening during higher-flux states, consistent with scenarios 

\onecolumngrid
\begingroup
\footnotesize
\setlength{\tabcolsep}{3pt}
\renewcommand{\arraystretch}{0.95}

\begin{longtable}{%
  >{\raggedright\arraybackslash}p{2.6cm} 
  >{\raggedright\arraybackslash}p{1.8cm} 
  >{\centering\arraybackslash}p{2.2cm}   
  >{\centering\arraybackslash}p{2.2cm}   
  >{\centering\arraybackslash}p{2.2cm}   
  >{\centering\arraybackslash}p{2.6cm}   
  >{\raggedright\arraybackslash}p{3.0cm} 
}

\caption{Spectral fit parameters for \textit{NICER} observations of Mrk~421: ObsID, Model (best-fitting highlighted in blue), Photon index $\Gamma$, $\alpha$, or $\Gamma_1$, Curvature $\beta$ or $\Gamma_2$, break/peak energy $E_b$ or $E_p$ (keV), reduced $\chi^2_{\mathrm{r}}$ (dof), and F-test value with probability values.}
\label{tab:spectral-params}
\\
\toprule
\textbf{Obs. ID} & \textbf{Model} &
\textbf{$\Gamma\,|\,\alpha\,|\,\Gamma_1$} &
\textbf{$\beta\,|\,\Gamma_2$} &
\textbf{$E_p\,|\,E_b$ (keV)} &
\textbf{$\chi^2_{\mathrm{r}}$ ($=\chi^2$/dof)} &

\textbf{F-value (prob.)} \\
\midrule
\endfirsthead

\caption[]{Spectral fit parameters for \textit{NICER} observations of Mrk~421 (continued).}
\\
\textbf{Obs. ID} & \textbf{Model} &
\textbf{$\Gamma\,|\,\alpha\,|\,\Gamma_1$} &
\textbf{$\beta\,|\,\Gamma_2$} &
\textbf{$E_p\,|\,E_b$ (keV)} &
\textbf{$\chi^2_{\mathrm{r}}$ ($=\chi^2$/dof)} &
 
\textbf{F-value (prob.)} \\
\midrule
\endhead

\midrule \multicolumn{7}{r}{Continued on next page}\\
\endfoot

\bottomrule
\endlastfoot
5100110101 & PL & $2.799 \pm 0.007$ & – & – & $7.29$ (867.2/119) & -- \\
& LP & $2.824 \pm 0.007$ & $0.453 \pm 0.024$ & $0.123 \pm 0.014$ & $5.23$ (617.7/118) & 47.7 ($2.72\times 10^{-10}$) \\
& \textcolor{blue}{BPL} & $2.858 \pm 0.008$ & $0.929 \pm 0.075$ & $3.034 \pm 0.054$ & $1.61$ (187.9/117) & 211.5 ($1.4\times 10^{-39}$) \\
5100110102 & PL & $2.865 \pm 0.007$ & – & – & $1.52$ (182.7/120) & -- \\
& LP & $2.857 \pm 0.007$ & $0.104 \pm 0.025$ & $7.6{\times}10^{-5}\,\pm\,1.8{\times}10^{-5}$ & $1.39$ (165.7/119) & 12.2 ($6.69\times 10^{-4}$) \\
& \textcolor{blue}{BPL} & $2.869 \pm 0.007$ & $0.699 \pm 0.427$ & $4.956 \pm 0.233$ & $1.26$ (148.2/118) & 13.7 ($4.34\times 10^{-6}$) \\
5100110104 & PL & $2.911 \pm 0.010$ & – & – & $1.41$ (147.1/104) & -- \\
& \textcolor{blue}{LP} & $2.910 \pm 0.010$ & $0.250 \pm 0.038$ & $0.015 \pm 0.010$ & $0.96$ (99.0/103) & 50.0 ($1.87\times 10^{-10}$) \\
& BPL & $2.907 \pm 0.010$ & $7.942 \pm 4.249$ & $5.381 \pm 0.417$ & $1.32$ (134.2/102) & 4.9 ($9.27\times 10^{-3}$) \\
6100110101 & PL & $2.071 \pm 0.003$ & – & – & $8.13$ (1137.8/140) & -- \\
& \textcolor{blue}{LP} & $1.964 \pm 0.005$ & $0.265 \pm 0.009$ & $1.169 \pm 0.026$ & $0.75$ (104.1/139) & 1380.3 ($4.64\times 10^{-74}$) \\
& BPL & $1.908 \pm 0.008$ & $2.226 \pm 0.007$ & $1.543 \pm 0.037$ & $1.07$ (147.9/138) & 461.8 ($7.23\times 10^{-62}$) \\
6100110102 & PL & $2.166 \pm 0.002$ & – & – & $13.16$(1974.0/150) & -- \\
& \textcolor{blue}{LP} & $2.029 \pm 0.004$ & $0.277 \pm 0.007$ & $0.886 \pm 0.015$ & $0.67$ (100.3/149) & 2783.5 ($2.63\times 10^{-98}$) \\
& BPL & $1.965 \pm 0.007$ & $2.315 \pm 0.005$ & $1.565 \pm 0.029$ & $1.03$ (152.5/148) & 883.9 ($5.08\times 10^{-83}$) \\
6100110103 & PL & $2.264 \pm 0.002$ & – & – & $11.81$(1735.4/147) & -- \\
& \textcolor{blue}{LP} & $2.136 \pm 0.004$ & $0.315 \pm 0.008$ & $0.608 \pm 0.012$ & $0.70$ (102.1/146) & 2335.6 ($1.03\times 10^{-91}$) \\
& BPL & $2.064 \pm 0.008$ & $2.437 \pm 0.006$ & $1.515 \pm 0.029$ & $0.96$ (138.9/145) & 833.3 ($3.09\times 10^{-80}$) \\
6100110104 & PL & $2.243 \pm 0.003$ & – & – & $10.05$(1377.3/137) & -- \\
& \textcolor{blue}{LP} & $2.127 \pm 0.005$ & $0.314 \pm 0.009$ & $0.628 \pm 0.014$ & $0.68$ (92.0/136) & 1900.0 ($8.57\times 10^{-82}$) \\
& BPL & $2.068 \pm 0.007$ & $2.429 \pm 0.008$ & $1.561 \pm 0.034$ & $1.42$ (192.0/135) & 416.7 ($1.73\times 10^{-58}$) \\
6100110110 & PL & $2.059 \pm 0.010$ & – & – & $2.07$ (217.3/105) & -- \\
& \textcolor{blue}{LP} & $2.001 \pm 0.012$ & $0.293 \pm 0.033$ & $0.996 \pm 0.047$ & $1.24$ (128.6/104) & 71.7 ($1.72\times 10^{-13}$) \\
& BPL & $2.024 \pm 0.012$ & $2.628 \pm 0.151$ & $3.362 \pm 0.294$ & $1.68$ (172.7/103) & 13.3 ($7.28\times 10^{-6}$) \\
6100110111 & PL & $2.027 \pm 0.003$ & – & – & $6.21$ (844.2/136) & -- \\
& \textcolor{blue}{LP} & $1.935 \pm 0.005$ & $0.256 \pm 0.010$ & $1.340 \pm 0.034$ & $0.76$ (102.6/135) & 975.8 ($1.21\times 10^{-63}$) \\
& BPL & $1.879 \pm 0.008$ & $2.182 \pm 0.008$ & $1.529 \pm 0.042$ & $0.86$ (115.7/134) & 421.9 ($1.48\times 10^{-58}$) \\
6100110116 & PL & $2.231 \pm 0.005$ & – & – & $3.47$ (416.5/120) & -- \\
& \textcolor{blue}{LP} & $2.164 \pm 0.007$ & $0.274 \pm 0.017$ & $0.502 \pm 0.026$ & $1.05$ (125.2/119) & 276.9 ($7.58\times 10^{-33}$) \\
& BPL & $2.096 \pm 0.012$ & $2.396 \pm 0.015$ & $1.370 \pm 0.056$ & $1.20$ (141.5/118) & 114.7 ($2.17\times 10^{-28}$) \\
6100110132 & PL & $2.365 \pm 0.004$ & – & – & $7.00$ (881.4/126) & -- \\
& \textcolor{blue}{LP}  & $2.254 \pm 0.006$ & $0.376 \pm 0.014$ & $0.459 \pm 0.016$ & $0.82$ (102.2/125) & 953.0 ($2.49\times 10^{-60}$) \\
& BPL & $2.176 \pm 0.011$ & $2.576 \pm 0.012$ & $1.420 \pm 0.035$ & $1.25$ (154.8/124) & 291.0 ($1.46\times 10^{-47}$) \\
6100110133 & PL & $2.459 \pm 0.004$ & – & – & $6.19$ (779.9/126) & -- \\
& \textcolor{blue}{LP}  & $2.359 \pm 0.006$ & $0.358 \pm 0.015$ & $0.315 \pm 0.016$ & $0.93$ (116.8/125) & 709.7 ($2.24\times 10^{-53}$) \\
& BPL & $2.290 \pm 0.010$ & $2.677 \pm 0.013$ & $1.471 \pm 0.038$ & $1.13$ (139.7/124) & 284.1 ($4.96\times 10^{-47}$) \\
6100110134 & PL & $2.475 \pm 0.003$ & – & – & $12.12$(1624.7/134) & -- \\
& \textcolor{blue}{LP} & $2.353 \pm 0.005$ & $0.363 \pm 0.010$ & $0.326 \pm 0.011$ & $0.72$ (96.3/133) & 2110.9 ($1.77\times 10^{-83}$) \\
& BPL & $2.266 \pm 0.008$ & $2.681 \pm 0.008$ & $1.447 \pm 0.028$ & $1.25$ (164.9/132) & 584.3 ($2.66\times 10^{-66}$) \\
6100110135 & PL & $2.486 \pm 0.003$ & – & – & $10.16$(1443.0/142) & -- \\
& \textcolor{blue}{LP} & $2.373 \pm 0.004$ & $0.306 \pm 0.009$ & $0.246 \pm 0.011$ & $0.98$ (137.9/141) & 1334.4 ($9.1\times 10^{-74}$) \\
& BPL & $2.274 \pm 0.009$ & $2.642 \pm 0.006$ & $1.347 \pm 0.028$ & $1.05$ (146.8/140) & 618.1 ($3.33\times 10^{-70}$) \\
6100110136 & PL & $2.469 \pm 0.003$ & – & – & $7.25$ (949.6/131) & -- \\
& \textcolor{blue}{LP} & $2.382 \pm 0.005$ & $0.320 \pm 0.012$ & $0.253 \pm 0.014$ & $0.92$ (119.1/130) & 906.5 ($1.85\times 10^{-60}$) \\
& BPL & $2.265 \pm 0.011$ & $2.632 \pm 0.008$ & $1.224 \pm 0.030$ & $1.02$ (131.1/129) & 402.7 ($3.42\times 10^{-56}$) \\
6100110137 & PL & $2.512 \pm 0.003$ & – & – & $12.09$(1740.4/144) & -- \\
& \textcolor{blue}{LP} & $2.388 \pm 0.004$ & $0.316 \pm 0.008$ & $0.243 \pm 0.009$ & $1.04$ (148.1/143) & 1537.5 ($2.14\times 10^{-78}$) \\
& BPL & $2.289 \pm 0.009$ & $2.674 \pm 0.006$ & $1.372 \pm 0.026$ & $1.31$ (185.4/142) & 595.5 ($8.91\times 10^{-70}$) \\
6100110141 & PL & $2.453 \pm 0.003$ & – & – & $9.40$ (1221.9/130) & -- \\
& \textcolor{blue}{LP} & $2.351 \pm 0.005$ & $0.367 \pm 0.012$ & $0.333 \pm 0.013$ & $0.84$ (107.7/129) & 1334.6 ($6.76\times 10^{-70}$) \\
& BPL & $2.248 \pm 0.009$ & $2.664 \pm 0.009$ & $1.362 \pm 0.029$ & $0.97$ (124.3/128) & 565.1 ($2.99\times 10^{-64}$) \\
6100110143 & PL & $2.614 \pm 0.004$ & – & – & $6.83$ (874.9/128) & -- \\
& \textcolor{blue}{LP} & $2.536 \pm 0.005$ & $0.357 \pm 0.013$ & $0.178 \pm 0.012$ & $0.71$ (90.3/127) & 1103.5 ($1.75\times 10^{-64}$) \\
& BPL & $2.428 \pm 0.010$ & $2.815 \pm 0.011$ & $1.292 \pm 0.032$ & $0.94$ (118.6/126) & 401.7 ($2.11\times 10^{-55}$) \\
6100110150 & PL & $2.482 \pm 0.003$ & – & – & $7.75$ (1015.5/131) & -- \\
& \textcolor{blue}{LP} & $2.392 \pm 0.005$ & $0.331 \pm 0.012$ & $0.256 \pm 0.013$ & $0.82$ (106.6/130) & 1108.4 ($1.73\times 10^{-65}$) \\
& BPL & $2.296 \pm 0.009$ & $2.665 \pm 0.009$ & $1.335 \pm 0.032$ & $1.13$ (146.3/129) & 383.2 ($5.34\times 10^{-55}$) \\
6704012801 & PL & $2.439 \pm 0.004$ & – & – & $6.52$ (841.0/129) & -- \\
& \textcolor{blue}{LP} & $2.356 \pm 0.005$ & $0.321 \pm 0.012$ & $0.279 \pm 0.014$ & $0.68$ (87.5/128) & 1102.3 ($9.39\times 10^{-65}$) \\
& BPL & $2.295 \pm 0.008$ & $2.652 \pm 0.012$ & $1.533 \pm 0.040$ & $1.11$ (140.6/127) & 316.3 ($4.7\times 10^{-50}$) \\
6704013201 & PL & $2.335 \pm 0.003$ & – & – & $9.10$ (1183.5/130) & -- \\
& \textcolor{blue}{LP} & $2.230 \pm 0.005$ & $0.360 \pm 0.012$ & $0.479 \pm 0.014$ & $0.64$ (83.2/129) & 1706.0 ($3.09\times 10^{-76}$) \\
& BPL & $2.138 \pm 0.009$ & $2.538 \pm 0.009$ & $1.381 \pm 0.031$ & $1.27$ (162.5/128) & 402.1 ($6.48\times 10^{-56}$) \\
6704013801 & PL & $2.307 \pm 0.003$ & – & – & $6.03$ (783.5/130) & -- \\
& \textcolor{blue}{LP} & $2.226 \pm 0.005$ & $0.275 \pm 0.011$ & $0.388 \pm 0.017$ & $0.97$ (124.7/129) & 681.5 ($2.52\times 10^{-53}$) \\
& BPL & $2.158 \pm 0.009$ & $2.470 \pm 0.009$ & $1.421 \pm 0.040$ & $1.21$ (154.8/128) & 259.9 ($8.45\times 10^{-46}$) \\
6704016701 & PL & $2.176 \pm 0.003$ & – & – & $5.49$ (729.5/133) & -- \\
& \textcolor{blue}{LP} & $2.095 \pm 0.005$ & $0.244 \pm 0.010$ & $0.639 \pm 0.019$ & $0.90$ (119.0/132) & 677.2 ($8.04\times 10^{-54}$) \\
& BPL & $2.021 \pm 0.009$ & $2.308 \pm 0.008$ & $1.348 \pm 0.040$ & $1.02$ (134.1/131) & 290.8 ($6.58\times 10^{-49}$) \\
6704016901 & PL & $2.170 \pm 0.003$ & – & – & $9.98$ (1387.0/139) & -- \\
& \textcolor{blue}{LP} & $2.052 \pm 0.005$ & $0.309 \pm 0.009$ & $0.824 \pm 0.016$ & $0.69$ (95.7/138) & 1862.1 ($5.32\times 10^{-82}$) \\
& BPL & $1.981 \pm 0.008$ & $2.344 \pm 0.007$ & $1.489 \pm 0.032$ & $1.13$ (155.1/137) & 544.1 ($6.68\times 10^{-66}$) \\
6704017001 & PL & $2.127 \pm 0.003$ & – & – & $6.94$ (964.1/139) & -- \\
& \textcolor{blue}{LP} & $2.031 \pm 0.005$ & $0.247 \pm 0.009$ & $0.865 \pm 0.021$ & $0.85$ (117.7/138) & 992.4 ($6.89\times 10^{-65}$) \\
& BPL & $1.969 \pm 0.008$ & $2.263 \pm 0.007$ & $1.458 \pm 0.039$ & $1.13$ (154.6/137) & 358.7 ($3.54\times 10^{-55}$) \\
6704017101 & PL & $2.164 \pm 0.003$ & – & – & $7.98$ (1117.7/140) & -- \\
& \textcolor{blue}{LP} & $2.060 \pm 0.004$ & $0.269 \pm 0.009$ & $0.774 \pm 0.015$ & $0.72$ (100.0/139) & 1414.6 ($9.8\times 10^{-75}$) \\
& BPL & $1.977 \pm 0.009$ & $2.303 \pm 0.006$ & $1.368 \pm 0.033$ & $0.97$ (133.7/138) & 507.8 ($2.34\times 10^{-64}$) \\
6704017201 & PL & $2.016 \pm 0.003$ & – & – & $14.72$(2060.6/140) & -- \\
& \textcolor{blue}{LP} & $1.868 \pm 0.005$ & $0.359 \pm 0.009$ & $1.527 \pm 0.029$ & $0.68$ (95.2/139) & 2869.6 ($1.08\times 10^{-94}$) \\
& BPL & $1.811 \pm 0.007$ & $2.236 \pm 0.008$ & $1.648 \pm 0.029$ & $1.58$ (218.6/138) & 581.4 ($5.89\times 10^{-68}$) \\
6704018501 & PL & $2.236 \pm 0.006$ & – & – & $2.72$ (294.2/108) & -- \\
& LP & $2.172 \pm 0.008$ & $0.277 \pm 0.021$ & $0.489 \pm 0.031$ & $0.98$ (105.3/107) & 191.9 ($1.28\times 10^{-25}$) \\
& \textcolor{blue}{BPL} & $2.045 \pm 0.021$ & $2.361 \pm 0.013$ & $1.083 \pm 0.050$ & $0.94$ (99.1/106) & 104.3 ($8.99\times 10^{-26}$) \\
6704020601 & PL & $2.283 \pm 0.003$ & – & – & $6.23$ (815.5/131) & -- \\
& \textcolor{blue}{LP} & $2.195 \pm 0.005$ & $0.287 \pm 0.011$ & $0.457 \pm 0.016$ & $0.58$ (75.2/130) & 1279.8 ($3.77\times 10^{-69}$) \\
& BPL & $2.150 \pm 0.008$ & $2.479 \pm 0.011$ & $1.630 \pm 0.044$ & $0.98$ (125.9/129) & 353.3 ($4.62\times 10^{-53}$) \\
6704028701 & PL & $2.454 \pm 0.003$ & – & – & $9.55$ (1251.3/131) & -- \\
& \textcolor{blue}{LP} & $2.352 \pm 0.005$ & $0.349 \pm 0.011$ & $0.313 \pm 0.013$ & $1.05$ (137.1/130) & 1056.5 ($2.8\times 10^{-64}$) \\
& BPL & $2.243 \pm 0.010$ & $2.644 \pm 0.008$ & $1.322 \pm 0.028$ & $1.11$ (142.6/129) & 501.5 ($1.45\times 10^{-61}$) \\
6704029501 & PL & $2.378 \pm 0.004$ & – & – & $7.50$ (974.7/130) & -- \\
& \textcolor{blue}{LP} & $2.285 \pm 0.005$ & $0.327 \pm 0.012$ & $0.367 \pm 0.015$ & $0.77$ (99.0/129) & 1141.1 ($6.38\times 10^{-66}$) \\
& BPL & $2.194 \pm 0.009$ & $2.564 \pm 0.009$ & $1.360 \pm 0.033$ & $1.00$ (127.5/128) & 425.3 ($2.92\times 10^{-57}$) \\
6704030801 & PL & $2.433 \pm 0.004$ & – & – & $6.70$ (858.0/128) & -- \\
& \textcolor{blue}{LP} & $2.332 \pm 0.006$ & $0.364 \pm 0.014$ & $0.350 \pm 0.016$ & $0.98$ (123.9/127) & 752.5 ($3.28\times 10^{-55}$) \\
& BPL & $2.268 \pm 0.010$ & $2.655 \pm 0.013$ & $1.492 \pm 0.038$ & $1.42$ (178.4/126) & 240.0 ($1.07\times 10^{-43}$) \\
6704031701 & PL & $2.374 \pm 0.003$ & – & – & $7.89$ (1041.4/132) & -- \\
& \textcolor{blue}{LP} & $2.279 \pm 0.005$ & $0.314 \pm 0.011$ & $0.360 \pm 0.014$ & $0.83$ (109.1/131) & 1119.4 ($4.89\times 10^{-66}$) \\
& BPL & $2.190 \pm 0.009$ & $2.549 \pm 0.008$ & $1.367 \pm 0.032$ & $1.03$ (133.3/130) & 442.8 ($9.31\times 10^{-59}$) \\
6704031901 & PL & $2.442 \pm 0.003$ & – & – & $7.98$ (1046.0/131) & -- \\
& \textcolor{blue}{LP} & $2.347 \pm 0.005$ & $0.336 \pm 0.011$ & $0.305 \pm 0.013$ & $0.67$ (87.7/130) & 1420.5 ($7.73\times 10^{-72}$) \\
& BPL & $2.267 \pm 0.009$ & $2.641 \pm 0.010$ & $1.423 \pm 0.033$ & $1.09$ (140.4/129) & 416.0 ($5.56\times 10^{-57}$) \\
6704032101 & PL & $2.602 \pm 0.004$ & – & – & $7.77$ (1002.3/129) & -- \\
& \textcolor{blue}{LP}& $2.518 \pm 0.005$ & $0.367 \pm 0.013$ & $0.197 \pm 0.012$ & $0.90$ (115.4/128) & 983.7 ($6.18\times 10^{-62}$) \\
& BPL & $2.402 \pm 0.010$ & $2.800 \pm 0.010$ & $1.263 \pm 0.030$ & $1.21$ (153.2/127) & 351.9 ($1.59\times 10^{-52}$) \\
6704032301 & PL & $2.739 \pm 0.005$ & – & – & $5.01$ (601.8/120) & -- \\
& \textcolor{blue}{LP} & $2.663 \pm 0.006$ & $0.383 \pm 0.018$ & $0.136 \pm 0.013$ & $0.92$ (108.9/119) & 538.6 ($5.39\times 10^{-46}$) \\
& BPL & $2.538 \pm 0.014$ & $2.924 \pm 0.013$ & $1.179 \pm 0.034$ & $1.18$ (139.2/118) & 196.1 ($3.07\times 10^{-38}$) \\
6704032801 & PL & $2.534 \pm 0.004$ & – & – & $7.07$ (898.0/127) & -- \\
& \textcolor{blue}{LP} & $2.447 \pm 0.005$ & $0.376 \pm 0.014$ & $0.254 \pm 0.014$ & $0.99$ (124.9/126) & 779.9 ($8.13\times 10^{-56}$) \\
& BPL & $2.321 \pm 0.012$ & $2.729 \pm 0.011$ & $1.228 \pm 0.030$ & $1.23$ (153.8/125) & 302.4 ($1.27\times 10^{-48}$) \\
6704033101 & PL & $2.300 \pm 0.003$ & – & – & $8.18$ (1087.8/133) & -- \\
& \textcolor{blue}{LP} & $2.201 \pm 0.005$ & $0.303 \pm 0.010$ & $0.466 \pm 0.015$ & $0.85$ (112.6/132) & 1143.2 ($7.14\times 10^{-67}$) \\
& BPL & $2.111 \pm 0.009$ & $2.464 \pm 0.008$ & $1.354 \pm 0.032$ & $1.12$ (146.4/131) & 421.2 ($8.89\times 10^{-58}$) \\
6704035701 & PL & $2.605 \pm 0.006$ & – & – & $3.02$ (350.3/116) & -- \\
& \textcolor{blue}{LP} & $2.551 \pm 0.007$ & $0.303 \pm 0.021$ & $0.123 \pm 0.018$ & $1.02$ (117.8/115) & 227.0 ($5.54\times 10^{-29}$) \\
& BPL & $2.458 \pm 0.015$ & $2.766 \pm 0.016$ & $1.217 \pm 0.051$ & $1.10$ (125.4/114) & 102.2 ($3.71\times 10^{-26}$) \\
6704035901 & PL & $2.455 \pm 0.005$ & – & – & $5.11$ (638.2/125) & -- \\
& \textcolor{blue}{LP} & $2.365 \pm 0.006$ & $0.339 \pm 0.016$ & $0.290 \pm 0.018$ & $0.92$ (113.7/124) & 572.0 ($2.79\times 10^{-48}$) \\
& BPL & $2.285 \pm 0.011$ & $2.642 \pm 0.012$ & $1.356 \pm 0.040$ & $1.18$ (145.2/123) & 208.8 ($2.86\times 10^{-40}$) \\
7100110102 & PL & $2.495 \pm 0.004$ & – & – & $6.09$ (761.2/125) & -- \\
& \textcolor{blue}{LP} & $2.397 \pm 0.006$ & $0.366 \pm 0.015$ & $0.287 \pm 0.016$ & $0.92$ (114.6/124) & 699.6 ($8.05\times 10^{-53}$) \\
& BPL & $2.329 \pm 0.010$ & $2.712 \pm 0.013$ & $1.452 \pm 0.039$ & $1.33$ (163.3/123) & 225.2 ($7.7\times 10^{-42}$) \\
7566013301 & PL & $2.493 \pm 0.006$ & – & – & $5.43$ (608.6/112) & -- \\
& \textcolor{blue}{LP} & $2.407 \pm 0.007$ & $0.421 \pm 0.021$ & $0.329 \pm 0.019$ & $1.19$ (131.7/111) & 401.9 ($1.09\times 10^{-38}$) \\
& BPL & $2.320 \pm 0.013$ & $2.761 \pm 0.020$ & $1.410 \pm 0.044$ & $1.62$ (177.8/110) & 133.3 ($4.05\times 10^{-30}$) \\
7566015901 & PL & $2.554 \pm 0.006$ & – & – & $2.99$ (346.3/116) & -- \\
& \textcolor{blue}{LP} & $2.497 \pm 0.007$ & $0.340 \pm 0.022$ & $0.186 \pm 0.021$ & $0.70$ (80.3/115) & 380.9 ($2.69\times 10^{-38}$) \\
& BPL & $2.367 \pm 0.018$ & $2.711 \pm 0.015$ & $1.097 \pm 0.043$ & $0.84$ (96.3/114) & 148.0 ($2.08\times 10^{-32}$) \\
7566016701 & PL & $2.535 \pm 0.005$ & – & – & $4.62$ (554.1/120) & -- \\
& \textcolor{blue}{LP} & $2.456 \pm 0.007$ & $0.383 \pm 0.019$ & $0.254 \pm 0.018$ & $0.85$ (101.1/119) & 533.2 ($8.82\times 10^{-46}$) \\
& BPL & $2.352 \pm 0.013$ & $2.749 \pm 0.015$ & $1.294 \pm 0.039$ & $0.93$ (109.2/118) & 240.4 ($2.42\times 10^{-42}$) \\
7566039501 & PL & $2.630 \pm 0.006$ & – & – & $4.26$ (494.3/116) & -- \\
& \textcolor{blue}{LP} & $2.560 \pm 0.007$ & $0.387 \pm 0.021$ & $0.189 \pm 0.018$ & $0.86$ (98.7/115) & 460.9 ($4.86\times 10^{-42}$) \\
& BPL & $2.465 \pm 0.013$ & $2.857 \pm 0.018$ & $1.323 \pm 0.043$ & $1.07$ (121.7/114) & 174.5 ($2.01\times 10^{-35}$) \\

\end{longtable}
\endgroup
\twocolumngrid
\noindent involving  enhanced particle acceleration or the dominance of higher-energy electrons in the emitting region.
Figure~\ref{fig:corr}(d) shows the correlation between the curvature parameter $\beta$ and the peak energy $E_p$ (keV).

The distribution exhibits an overall negative trend ($r = -0.52$), with $\beta$ decreasing as $E_p$ increases, particularly pronounced for $E_p \gtrsim 0.5$ keV. At lower $E_p$ values, the data show significant scatter, indicating increased dispersion in $\beta$. Although the general trend suggests an inverse power law dependence of $\beta$ on $E_p$, the observed spread implies that additional physical parameters may contribute to the dispersion. We investigated correlations between the log-parabolic photon index $\alpha$ and both the flux and synchrotron peak energy $E_{\mathrm{p}}$. 

Figure~\ref{fig:corr}(e) shows a strong negative correlation between $\alpha$ and $E_{\mathrm{p}}$, with $r=-0.88$ (Pearson).~An inverse power-law model provides an excellent description of the trend, indicating that spectra with higher synchrotron peak energies systematically exhibit smaller (harder) values of $\alpha$.

Similarly, Figure~\ref{fig:corr}(f) shows a clear negative correlation  between $\alpha$ and flux, with a Pearson correlation coefficient of $r=-0.83$. This trend indicates that the X-ray spectrum becomes flatter as the flux increases.
\section{Discussion}\label{sec:discussion}
During the two-year \textit{NICER} monitoring campaign, Mrk 421 exhibited pronounced long-term X-ray variability, with the average flux increasing by a factor of $\sim$28 and a total fractional variability amplitude of $\sim$48\%.  
The source reached its most active state on 2023 May 24 and a comparatively quiescent state on 2024 March 19. On intra-day timescales, the source showed rapid variability, showing a fractional variability of 
$\sim$8\%  within a span of $\sim$ 40 min.

\subsection{Constraints on the Emission Region and Magnetic Field}

The rapid X-ray variability observed in Mrk~421 provides important constraints on the physical conditions within the emitting region. In high-synchrotron-peaked blazars, the X-ray emission is generally interpreted as synchrotron radiation produced by the highest-energy electrons in the jet. Therefore, the observed variability timescales can be directly associated with synchrotron cooling processes.
Assuming that the observed rapid variability is governed by synchrotron cooling, and it can be expressed as,

\begin{equation}
t_{\rm cool}
=
\frac{3}{4}
\frac{m_e c}
{\sigma_T U_B \gamma \beta^2}
\sim
7.74 \times 10^{8}
\gamma^{-1} B^{-2}
\ {\rm s},
\tag{13}
\end{equation}
where $\gamma$ is the Lorentz factor of the emitting electrons, $B$ is the magnetic field strength in Gauss, $\sigma_T$ is the Thomson cross-section, and $U_B$ is the magnetic energy density. For ultra-relativistic electrons ($\beta \sim 1$), the magnetic field corresponding to the synchrotron cooling timescale can be estimated as,

\begin{equation}
B=
\frac{
2.09 \times 10^{2}
(1+z)^{1/3}
}{
t_{\rm cool}^{2/3}
\delta^{1/3}
E^{1/3}
}
\tag{14}
\end{equation}
where $E$ is the observed photon energy in keV, $\delta$ is the Doppler factor, and $z$ is the source redshift.
In the case of Mrk~421 ($z = 0.031$), assuming that the observed variability timescale represents the synchrotron cooling timescale, with $t_{\rm var} \sim t_{\rm cool} \sim 40$ min, along with a typical Doppler factor $\delta = 10$ and an X-ray photon energy of $E \sim 1$ keV, we estimate the magnetic field strength to be approximately $B \sim 0.5~{\rm G}$.

Based on the causality arguments, the observed rapid variability places an upper limit on the size of the emitting region. Using the relation, 

\begin{equation}
R \lesssim
\frac{
\delta
}{
1+z
}
c\, t_{\rm var}
\tag{15}
\end{equation}
for a typical $\delta = 10$ and $z = 0.031$, the size of the emitting region can be estimated to be
$R \lesssim 7.0 \times 10^{14}\ {\rm cm}$, which suggests that the X-ray emission originates from a highly compact region within the relativistic jet. Furthermore, assuming a conical jet geometry with an opening angle of $\theta_j \sim 1/\Gamma$ and a bulk Lorentz factor of $\Gamma \sim 10$, the distance of the emitting region from the central black hole can be estimated as $d \sim R/\theta_j \sim 7 \times 10^{15}\ {\rm cm}$. For a typical AGN black hole mass of $10^9\,M_{\odot}$, this corresponds to a distance of a few hundred Schwarzschild radii, where the Schwarzschild radius is given by $R_{\rm s} = 2GM/c^{2}$.
These results imply that the observed X-ray emission originates from a compact region located close to the central engine, where efficient synchrotron cooling processes are actively operating.

The observed large-amplitude variations and rapid variations are characteristic of high-synchrotron-peaked blazars and are generally attributed to fluctuations in particle acceleration efficiency and radiative cooling within the jet \citep{fossati2000x, khatoon2022correlations, Aleksic2015, Markowitz2022, bhatta2018hard}.
Variability across different flux states was examined through the correlation between RMS and mean flux; however, no clear RMS–flux relation was detected in our analysis of Mrk 421. We note that some studies do report a definite correlation. For example, \citet{bhattacharyya2020blazar} found a linear trend using multi-mission data from \textit{XMM-Newton}, \textit{AstroSat}, \textit{ASCA}, and \textit{Swift}/XRT, suggesting that such behavior may emerge on longer timescales. More recently, \citet{Bhatta2025Nustar} reported a power-law–dependent RMS–flux relation in \textit{NuSTAR} observations of the source.~However, the lack of a clear RMS–flux relation and the multi-peaked distribution observed in our \textit{NICER} data may result from limited sampling and irregular cadence, which can bias the sampled flux range, as well as from the instrument’s ability to resolve faint and bright states with high sensitivity \citep[see also,][]{bhatta2025probing,dinesh2023constraining}. 

Previous analyses of Mrk~421 with \textit{Swift/XRT} have shown that its flux distribution is well described by a log-normal profile \citep{kapanadze2016}, and similar behavior has also been reported in \textit{XMM-Newton} observations of other blazars \citep{Wani2022}. However, the flux distribution derived from the complete observation set shows significant deviations from both Gaussian and lognormal forms. Rather than a simple unimodal profile, the histogram reveals a complex, multimodal shape. Such results suggest that the observed variability may reflect the combined contributions of multiple emission zones, each subject to distinct physical conditions, including variations in particle acceleration efficiency, magnetic field strength, and Doppler boosting within the jet, indicating that the variability cannot be fully described by a single stochastic process.

In the model independent method of spectral analysis HR were found to be significantly correlated with flux states. The correlation exhibits a pronounced \textit{harder-when-brighter} behavior, in agreement with previous studies \citep{Tramacere2009,kapanadze2018swift}. 
Similarly, a strong correlation between the high- and low-energy count rates, reported across multiple X-ray instruments including \textit{XMM-Newton, Swift/XRT, Chandra, and NuSTAR} \citep{Aggrawal2018, kapanadze2018swift, Gokus2024, Noel2022, Pandey2017}, indicates that the variability in both bands is closely linked, implying a common origin of the X-ray emission within the jet.

In our investigation of the spectral shape of the source across multiple flux states, we performed spectral modeling using three widely adopted models: a power law, a broken power law, and a log-parabola. Our results show that, for the majority of the epochs, the spectral shape is better represented by a log-parabolic model, with mean $\alpha \approx 2.32 \pm 0.234$ and curvature parameter $\beta \approx 0.32 \pm 0.058$, indicating the presence of strong spectral curvature. Moreover,  we examined correlations among the key spectral parameters derived from the spectral models. 
We find that the log-parabolic spectral parameters \(\alpha\) and \(\beta\) are positively correlated, with a Pearson coefficient of \(r = 0.72\). Similar correlation studies have also been reported in earlier investigations, highlighting the importance of such analyses, and such a correlation is generally interpreted as a signature of first-order Fermi acceleration with an energy-dependent acceleration probability \citep{Wani2023,kapanadze2018swift,massaro2004log}. 


Similarly, the analysis results that $\beta$ found to be negatively correlated with $E_{\rm p}$, in the sense that the curvature parameter gets smaller when the synchrotron peak shifts towards higher energy. Such an \(E_{\mathrm{p}}-\beta\) anti-correlation has also been reported in several similar studies of TeV blazars based on observations from various instruments, including \textit{Swift, NuSTAR, and AstroSat} \citep[e.g.,][]{das2025sub,kapandze2017,kapanadze2016,Tramacere2009,tramacere2007signatures,massaro2008,massaro2006}, suggesting that this behavior may be a common characteristic of synchrotron-dominated blazars. The correlation can be given within the framework of statistical acceleration \citep{massaro2006}, and often interpreted within two theoretical frameworks: the energy-dependent acceleration probability (EDAP) scenario, where acceleration efficiency decreases with particle energy, and the stochastic acceleration (SA) framework \citep{Tramacere2009}. In the context of EDAP scenario, such an inverse dependence of acceleration efficiency on particle energy naturally produces a log-parabolic particle energy distribution in which the spectral curvature \(r\) is related to the fractional acceleration gain \(\epsilon\) by \(r \propto 1 / \log \epsilon\) \citep{massaro2004log}.

Similarly, the observed anti-correlation can also  in the SA framework  turbulence-driven random processes energize particles, and the momentum-diffusion term broadens the electron distribution, in such a way that  the curvature parameter \(r\) scales inversely with the product \(D\,t\), where \(D\) is the momentum diffusion coefficient and \(t\) is the acceleration timescale. This scaling establishes a direct connection between the electron peak energy and the synchrotron curvature, with the photon spectral curvature related to the particle distribution as \(\beta \simeq r/4\) \citep{massaro2004log, tramacere2007signatures, Tramacere2011stochastic}. Moreover, in the blazar jets the confinement of charged particles within the acceleration region depends on their gyroradius; as the Larmor radius $R_{\mathrm{L}}$ grows, particles become less confined, and the probability of acceleration, $P_{\mathrm{acc}}$, correspondingly decreases. From the relations \(E_{\mathrm{p}} \propto \epsilon\) and \(r \propto 1 / \log \epsilon\), a negative correlation \(E_{\mathrm{p}}-\beta\) emerges, in agreement with the observed trend.

A consistent \textit{harder-when-brighter} behavior is evident in our results. We find that the synchrotron peak energy \(E_{\mathrm{p}}\) increases with X-ray flux, indicating spectral hardening as the source brightens, in agreement with earlier studies of Mrk~421 \citep{tanihata2004evolution, kapanadze2018swift}. 

This behavior is further supported by the observed anti-correlations between \(\alpha\) (or equivalently \(\Gamma\)) and both \(E_{\mathrm{peak}}\) and the total flux: the spectrum becomes flatter (smaller \(\alpha\)) when the peak shifts to higher energies and when the source enters a brighter state. Similar harder-when-brighter trends have been reported in previous studies \citep{das2025sub, Wani2023, Hota2021}.
\vspace{0.5cm}
\subsection{Role of Curvature Parameter in Shaping the Spectral Properties}

To investigate the effect of spectral curvature on the synchrotron emission, we compute model spectra using the log-parabola (LP) model for different values of the curvature parameter ($r$). In this model, the underlying electron energy distribution is given by,

\begin{equation}
N(\gamma) \propto \gamma^{-p - r \log_{10}(\gamma)}
\tag{16}
\end{equation}
where $p$ is the electron spectral index and $r$ is the curvature parameter. This form naturally describes curved spectra and is commonly associated with stochastic, energy-dependent acceleration processes in high-synchrotron-peaked blazars such as Mrk 421.
\begin{figure}[t]
    \centering
    \includegraphics[width=0.95\columnwidth]{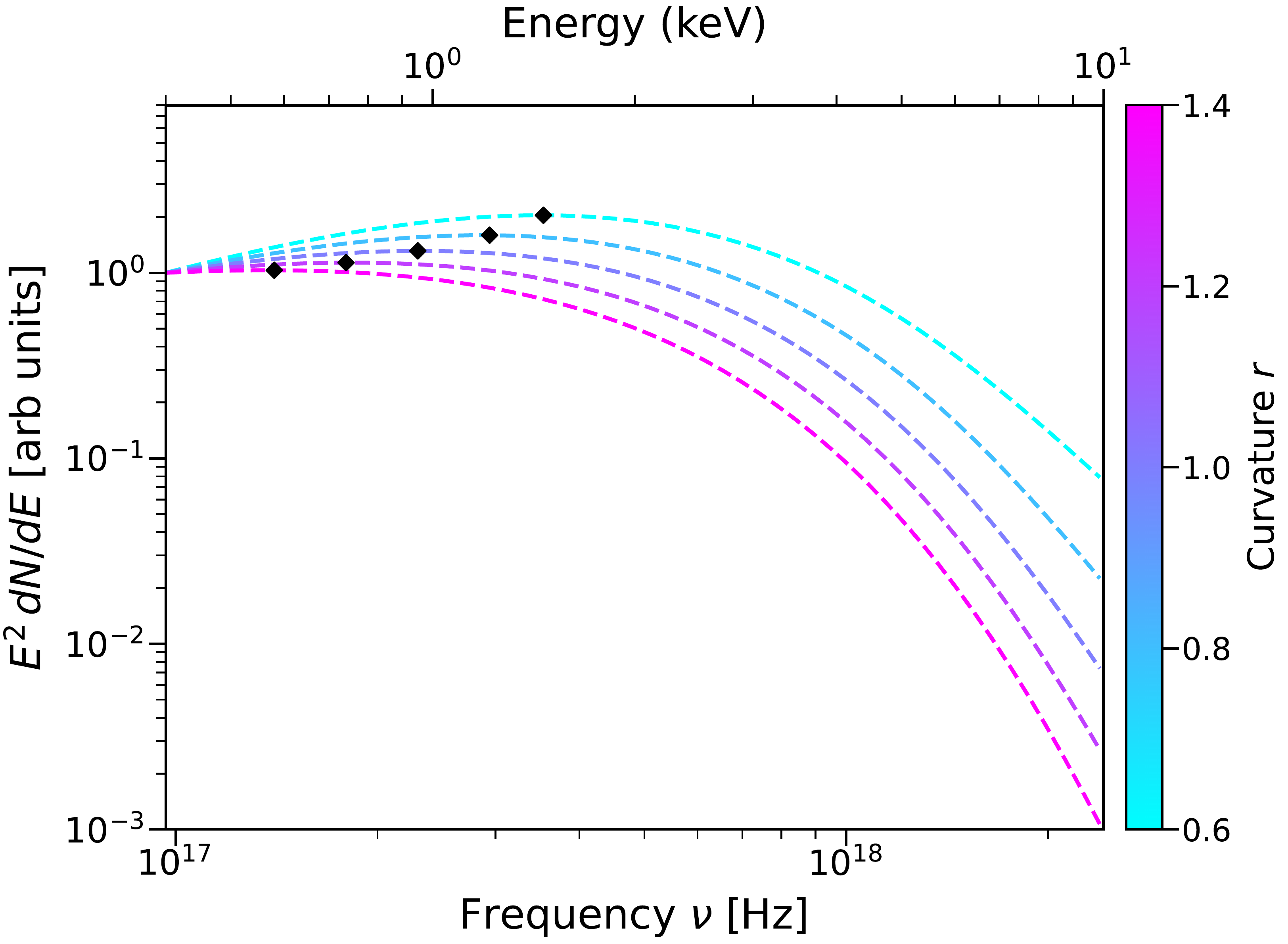}
    \caption{Synchrotron spectral energy distributions for different curvature parameters, $r$.
The spectra are normalized at 0.4 keV for better comparison. It is seen that larger $r$ values produce stronger curvature and lower peak energies. The black diamond points indicate the spectral peaks.
}
    \label{fig:vary_curvature}
\end{figure}
The synchrotron emission spectra ($j_{\nu}$) is obtained by integrating the electron energy distribution, $N(\gamma)$, with the single-particle synchrotron power,

\begin{equation}
j_{\nu} = \int P(\nu,\gamma)\, N(\gamma)\, d\gamma
\label{eq:synch_emissivity}
\tag{17}
\end{equation}
where $P(\nu,\gamma)$ represents the power spectrum emitted by a single electron, which can be approximated by,

\begin{equation}
P(\nu,\gamma) \propto z^{0.3} \exp(-z)
\tag{18}
\end{equation}
where

\[
z = \frac{4\pi \nu}{3 \, \Omega \, \langle \sin\theta \rangle \, \gamma^{2}}
\]
and

\[
\Omega = \frac{eB}{m_e c}
\]
represents the cyclotron frequency and the angle between the magnetic field direction and the line of sight, $\theta$, is assumed to be $90^\circ$ for simplicity \citep[see][]{kirk1998particle}.

The synchrotron spectrum is obtained by numerically integrating Eq.~(\ref{eq:synch_emissivity}) for fixed values of $B = 0.5$ G and $p = 3.8$. The resulting spectral energy distributions for different values of $r$ are shown in Figure~\ref{fig:vary_curvature}.

We also explore the relation between the synchrotron peak energy ($E_p$) and the curvature parameter ($r$), and find a clear anti-correlation in which $E_p$ decreases with increasing $r$. This trend indicates that higher curvature corresponds to a stronger suppression of high-energy electrons, resulting in a systematic shift of synchrotron peak toward lower energies. Such a trend is in line with models in which the curvature of the particle distribution arises from energy-dependent acceleration efficiencies or radiative cooling. This behavior is also in agreement with our observational results, and the observed trend suggests that spectral curvature can serve as an important diagnostic of the underlying particle acceleration mechanism operating within the jet.

\vspace{0.5cm}
\section{Conclusion}\label{sec:conclusion}
In this study, we investigated the X-ray properties of the TeV blazar Mrk 421, focusing on its intra-day and long-term spectral and timing variability. We utilized two years of archival observations from the \textit{NICER} telescope and performed a detailed spectral and timing analysis, the main conclusions of which are summarized as follows.

\begin{itemize}
    \item  Mrk~421 is a highly variable source exhibiting fluctuations on multiple timescales and shows a total fractional variability of approximately 48\% across all \textit{NICER} observations.
    \item The multimodal count-rate distribution of Mrk~421, characterized by three distinct flux states, suggests that the observed variability cannot be explained by a single stochastic process and instead indicates the presence of multiple emission states within the jet.
    \item  The hardness ratio (HR) analysis reveals that the X-ray spectrum becomes harder as the source brightens.~This distribution confirms a pronounced \textit{harder-when-brighter} behavior.
    \item The correlation between the low-energy (LE) and high-energy (HE) count rates is strong and positive. The power-law fit yields an index of $\sim$1.17, indicating that the HE flux rises slightly faster than the LE flux. This behavior suggests that variability in both bands is tightly coupled and likely produced by the same population of relativistic electrons.
    \item The fractional variability analysis in the soft X-ray bands (0.4 to 1 keV) and the hard bands (1 to 10 keV) shows a near-linear relationship between the two, closely following the trend $y=x$.
    \item Our results confirm that, given adequate statistics, the X-ray spectra of Mrk~421 are best characterized by a log-parabolic model confirmed by the F-test.
    \item Our spectral analysis further reveals several systematic correlations among the spectral parameters and derived quantities. The log-parabolic parameters \(\alpha\) and \(\beta\) are positively correlated, while \(\beta\) is negatively correlated with the synchrotron peak energy \(E_{\mathrm{p}}\). 
    \item We also find that \(E_{\mathrm{p}}\) is positively correlated with the flux and negatively correlated with \(\alpha\). In addition, both \(\alpha\) and the photon index show negative correlations with flux, reflecting the well-known \textit{harder-when-brighter} behavior of the source.
    \item Using a log-parabolic synchrotron model, we find that increasing spectral curvature shifts the synchrotron peak toward lower energies, reproducing the observed anti-correlation between $E_{\rm p}$ and $\beta$ and supporting stochastic or energy-dependent particle acceleration scenarios.

\end{itemize}

In conclusion, this work presents the first detailed spectral and timing analysis of Mrk~421 using \textit{NICER} data. The results demonstrate that the source exhibits strong variability across different timescales and that its X-ray spectra are consistently best represented by a log-parabolic model.  Our analysis reveals a clear \textit{harder-when-brighter} trend, as well as systematic correlations between spectral parameters, flux, and synchrotron peak energy. We also show that synchrotron spectra generated using a log-parabolic electron energy distribution naturally reproduce the observed spectral evolution and the anti-correlation between $E_{\rm p}$ and $\beta$.~These findings highlight the role of energy-dependent particle acceleration and cooling in shaping the observed emission and motivate future work by combining \textit{NICER} data with multiwavelength campaigns to better understand the dynamics of relativistic jets.

\section{Acknowledgments}
We are grateful to the anonymous referee for the careful assessment and constructive suggestions, which have greatly improved the clarity and presentation of this paper.
This work was partially supported by
a program of the Polish Ministry of Science under the title “Regional Excellence Initiative”, project No. RID/SP/0050/2024/1.
TPA acknowledges the support of the National Natural Science 
Foundation of China (grant nos.\ 12222304, 12192220, and 12192221).

\bibliography{sangeetha}{}
\bibliographystyle{aasjournal}



\end{document}